\documentclass[aps,pre,preprint,groupedaddress,
,superscriptaddress,showpacs,a4paper,floatfix]{revtex4-1}

\usepackage{amsmath,amssymb,graphicx,graphics,bm}
\usepackage{hyperref}
\usepackage{graphics,graphicx}
\usepackage{amsmath,amssymb}
\usepackage{color}

\begin{document}

\title{Brownian motion: non-equilibrium states from equilibrium trajectories
-
recovering hydrodynamic regimes from prepared displacement measurements}

\author{Jason Boynewicz}
\affiliation{Department of Physics, The University of Texas at Austin, Austin, TX 78712, USA}
\author{Michael C. Thumann}
\affiliation{Department of Physics, The University of Texas at Austin, Austin, TX 78712, USA}
\author{Giuseppe Procopio}
\affiliation{Facolt\`{a} di Ingegneria Civile and Industriale, DICMA,  Universit\`{a} di Roma La Sapienza, via Eudossiana 18, 00184, Roma, Italy}
\author{Massimiliano Giona}
\email[corresponding author:]{massimiliano.giona@uniroma1.it}
\affiliation{Facolt\`{a} di Ingegneria Civile and Industriale, DICMA, Universit\`{a} di Roma La Sapienza, via Eudossiana 18, 00184, Roma, Italy}

\date{\today}

\begin{abstract}
Owing to the Chapman-Kolmogorov equation for Markovian dynamics,
any equilibrium trajectory of a Brownian particle in a solvent
fluid can be viewed as the superposition of an
uncountable number of non-equilibrium  states. This
property permits the unraveling of fine details of fluid-particle
interactions at microscales  defined by its non-equilibrium
properties from the analysis of a single Brownian trajectory
and to connect them to the hydrodynamics  of the solvent fluid,
simply considering the lower-order (second) moments of particle position
in trapped conditions. In this way, the acceleration due to 
thermal-hydrodynamic fluctuational forces is isolated from the
other factors and the short-time displacement statistics
is completely determined  by the correlation properties
of the fluctuational thermal-hydrodynamic force.
This approach not only confirms  
 the $t^{5/2}$-law obtained by Boynewicz et al. (2026),
related to fluid inertial effects, but indicates that
this scaling may be superseded by a $t^4$-scaling at very short times
once the correlated nature of the stochastic forcings is taken into
account. The latter result is related to the regularity properties
of particle velocity realizations. 
\end{abstract}

\maketitle

\section{Introduction}
\label{sec1}

Brownian motion plays a central role in the whole
spectrum of physical investigations  well beyond the sheer
phenomenology of equilibrium thermal motion of micrometric
rigid particles in a solvent fluid \cite{chandra}. From the diffusive
analysis of Brownian trajectories, the atomistic view
was definitively established by Einstein in 1905-06 \cite{einstein1,einstein2}, and the stochastic
approach put forward by Langevin in 1908 \cite{langevin} has been the
triggering point for the massive development, formalization
and use of stochastic differential equations in statistical
physical and more generally in all the branches of quantitative
sciences (biology, social sciences, etc.) \cite{coffey,vankampen}.
The use of Brownian particles as probes for unveiling experimentally
physical phenomena at microscale and below is the conceptual principle
underlying microrheology \cite{microrhe1,microrhe2} 
with implications in soft matter  sciences, biological systems,
polymer physics, and hydrodynamics.

It is therefore not surprising that the breakthrough of
highly resolved measurements of Brownian trajectories performed in the
last two decades \cite{hres1,hres2,hres3,hres4, madsen_ultrafast_2021}
have raised great expectations not only for
unveiling the fine structure of Brownian dynamics \cite{deemed}, but also
for the possibility of deriving new fundamental and novel
results in microphysics, statistical thermodynamics, and hydrodynamics.
While from one hand, the experimental analysis has confirmed the
results of hydrodynamic theory,  especially regarding the
influence  of the inertial effects (Basset force) at short time scales
\cite{hres2,hres3},
it has also raised new questions of fundamental nature such as added-mass effects and its implications
for the foundations of statistical physics \cite{opticsexp,gpp_pof}.

Most of the recent experiments on Brownian  motion refer  to
equilibrium trajectories and to the analysis of
equilibrium second-order moments (for particle
position and velocity), equilibrium densities, and velocity
autocorrelation functions. Very recently the analysis of non-equilibrium
states and their temporal propagation obtained from  the subsampling of
equilibrium trajectories has been performed showing the
occurrence of new phenomena associated with 
fluid inertial effects at short time scales \cite{btr}. This
work is the  extension to liquid media of
 the analysis previously developed by Duplat et al.  \cite{duplat}
for Brownian particles in gases and by Wang and Masoliver \cite{wm}
for generic Langevin equations in the presence of thermal or external
colored noise.
Beyond the specific results regarding hydrodynamical Brownian motion,
these  contributions have paved the way towards the experimental
analysis of non-equilibrium states from the sampling of
equilibrium trajectories.

The  aim of the present article is to develop the theory
underlying this approach and to derive its hydromechanical implications
in terms of the fine structure of Brownian dynamics.
This is because the short-time non-equilibrium  scaling
of the mean square displacement provides a fairly complete
characterization of the hydrodynamic regimes and of the
regularity properties of Brownian particle velocity.
  Indeed, the
possibility of extracting non-equilibrium dynamics from
equilibrium trajectories is imbedded in the Chapman-Kolmogorov equation
that represents the foundative relation of Markovian dynamics,
and can be extended to the non-Markovian case, upon
the establishment of a suitable Markovian embedding.

Particular attention is focused on the analysis of the second-order
moments
of the particle position $m_{xx}(t) = \langle x^2(t) \rangle$ starting
from specific preparations of the initial condition obtained via
a  suitable subsampling of the equilibrium trajectories.
The initial scaling of $m_{xx}(t)$ in specific and experimentally
accessible initial settings,  provides an accurate qualitative way 
to assess the relevance of the different hydrodynamic and rheological
effects (viscoelasticity, fluid inertial effects, etc.), ultimately
addressing the issue of the regularity (in the sense of Lipschitz/H\"older
continuity) of Brownian dynamics.  At the dynamic core of Brownian
motion is momentum transfer in the  fluid-particle interactions
at thermal equilibrium, and therefore the regularity of Brownian dynamics
refers to local properties of realizations of the
particle velocity. In fact, there is a direct connection between the
scaling exponent of $m_{xx}(t)$ at short time scales with the
regularity H\"older exponent of the particle velocity realizations.

The article is organized as follows. 
Section \ref{sec2} addresses the foundations of the extraction
of non-equilibrium states/dynamics from equilibrium trajectories via
specific subsamplings of the latter both in the Markovian
and in the non-Markovian cases. The starting point is
the Chapman-Kolmogorov equation,  indicating that any equilibrium state
or trajectory is the superposition of all the possible
non-equilibrium states (trajectories) weighted with respect to
the equilibrium density. 
Section \ref{sec3} reviews Brownian motion
hydrodynamics, expressed either
in the form of a Generalized Langevin Equation (GLE)
or  via   a Markovian embedding (modal approach) \cite{gpp1},
in the presence of generic hydrodynamic memory kernels  associated
either with viscoelastic dissipative effects or fluid inertial
(Basset force) contributions. The concept of
fluctuational
patterns \cite{gpp1}, consistent
with the Kubo fluctuation-dissipation theory is also reviewed.
 Section \ref{sec4} develops the analysis of
the initial scaling of $m_{xx}(t)$ and its relation with hydrodynamic
properties. 
The analysis is first grounded on the GLE formulation
of Brownian dynamics and subsequently
extended  to the modal representation. The modal representation clarifies some physical shortcomings of the 
GLE approach in the presence of inertial effects
(see Appendix \ref{app1}) that are of conceptual relevance but 
inconsequential in the analysis of short time properties.
All the hydrodynamic and transport conditions of physical
interest, including also anomalous
diffusion arising either from a long-term power-law scaling
of the memory kernels or from singular constrained geometries
(Brownian motion in a fractal structure), are considered.
To  focus on the physical results, the detailed analysis
of these cases is deferred to Appendix \ref{app2}.
While the $t^{5/2}$-law derived in \cite{btr} is recovered
in the presence of the singular Basset kernel, the occurrence 
of correlated fluctuations (in the meaning of correlated
fluctuational patterns introduced in \cite{gpp3}) determines an earlier $t^4$-scaling, and
a crossover between these two regimes.
The occurrence of the latter scaling is related to the regularity of
the particle velocity realizations, and this is thoroughly
analyzed in Section \ref{sec5}. Specifically, the scaling
of $m_{xx}(t)$ in the preparation where both the particle
position and velocity are vanishing (the origin of the laboratory
reference system is set 
at the position corresponding to the equilibrium point of
theoretical trap) can be related to the H\"older exponent of
the particle velocity and ultimately to its fractal dimension.
These results are confirmed by recent experiments \cite{btr,noifrac1}.

\section{Equilibrium trajectories and non-equilibrium states}
\label{sec2}

In order to illustrate the formal relation between
the equilibrium trajectories and non-equilibrium states,
consider the case of a generic Markovian process ${\bf X}(t)$
(we use the upper case lettering for indicating the process,
and lower case one, say ${\bf x}(t)$, for a generic realization of it).

For a $n$-dimensional vector-valued Markovian process ${\bf X}(t)$ the Chapman-Kolmogorov
equation applies: if $p({\bf x},t)$ is the probability density
for ${\bf X}(t)$ at time $t$, and $p({\bf x} , t \, | \, {\bf x}_0 , t_0)$
 the conditional (transition) probability  of having ${\bf X}(t)={\bf x}$ at time
$t$ if ${\bf X}(t_0)={\bf x}_0$ at $t_0<t$, the Chapman-Kolmogorov equation
reads
\begin{equation}
p({\bf x},t)= \int_{{\mathbb R}^n} p({\bf x} , t \, | \, {\bf x}_0 , t_0)
\, p({\bf x}_0,  t_0) \, d {\bf x}_0  \, ,
\label{eq3_1}
\end{equation}
where  $d {\bf x}_0= d x_{0,1} \cdots d x_{0,n}$ indicates the $n$-dimensional
measure element with respect to the ${\bf x}_0$-coordinates.

Suppose that for $t \rightarrow \infty$ an equilibrium state
is established, and let $p^*({\bf x})$ be the corresponding
equilibrium probability density.
Applying eq. (\ref{eq3_1}) for $t_0=0$ starting from the
equilibrium distribution,  i.e. $p({\bf x}_0,0)=p^*({\bf x}_0)$, and
thus $p({\bf x},t)=p^*({\bf x})$ for any $t>t_0=0$, we obtain the
following representation of the equilibrium distribution
\begin{equation}
p^*({\bf x})= \int_{{\mathbb R}^n} p({\bf x}, t \, | \, {\bf x}_0, 0)
\, p^*({\bf x}_0) \, d {\bf x}_0  \, .
\label{eq3_2}
\end{equation}
The latter representation of equilibrium densities admits
an interesting interpretation: the equilibrium state can
be viewed as the linear combination of non-equilibrium
conditions expressed by the conditional transition term
$p({\bf x}, t \, | \, {\bf x}_0, 0)$ corresponding to the
initial preparation of the system associated with
the initial density $p({\bf x},t=0)=\delta({\bf x}-{\bf x}_0)$
of impulsive nature centered at ${\bf x}={\bf x}_0$ and
weighted with respect to the equilibrium distribution  itself.
This result is not surprising as the equilibrium distribution $p^*({\bf x})$
is always the dominant (Frobenius) eigenfunction of
the  Fokker-Planck operator associated with the particle density
dynamics, admitting the constant function as its dual eigenfunction
counterpart (i.e. as the dominant eigenfunction of the
adjoint Fokker-Planck operator), while the
remaining eigenfunctions of the spectrum possess eigenvalues
with negative real part, as an equilibrium distribution is supposed
to exist and be stable.

Equation (\ref{eq3_2}) can be regarded as a path-integral
formulation of the equilibrium state: the equilibrium state is
simply the sum (integral) of all the non-equilibrium
paths originating from generic initial conditions ${\bf x}_0$, the
frequency of occurrence of which is simply proportional to the
equilibrium density  itself.

The same approach can be extended to the non-Markovian case
if a Markovian embedding of the stochastic particle dynamics exists.
Let us assume that ${\bf X}(t)$ is non-Markovian, but there exists
a system of $n_h$  stochastic processes ${\bf Q}(t)$ such
that the couple $({\bf X}(t),{\bf Q}(t))$ represents a $n+n_h$ Markovian
embedding of ${\bf X}(t)$.
In this case, letting $p^*({\bf x},{\bf q})$
be the equilibrium distribution, eq. (\ref{eq3_2}) is transformed
into
\begin{equation}
p^*({\bf x},{\bf q})= \int_{{\mathbb R}^n} d {\bf x}_0
\int_{{\mathbb R}^{n_h}} p({\bf x},{\bf q}, t \, | \, {\bf x}_0, {\bf q}_0, 0)
\, p^*({\bf x}_0 , {\bf q}_0) \, d {\bf q}_0  \, .
\label{eq3_3}
\end{equation}
Since $p^*({\bf x}_0,{\bf q}_0)= p^*({\bf q}_0 \, | \, {\bf x}_0) \, p_x^*({\bf x}_0)$,
the integration of both left and right hand sides of eq.
(\ref{eq3_3}) provides the expression of the equilibrium marginal density
$p_x^*({\bf x})$ of ${\bf x}$
\begin{eqnarray}
p_x^*({\bf x}) & = & \int_{{\mathbb R}^{n_h}} d{\bf q} \int_{{\mathbb R}^n}
d {\bf x}_0 \int_{{\mathbb R}^{n_h}} p({\bf x},{\bf q}, t \, | \, {\bf x}_0, {\bf q}_0, 0)
p^*({\bf q}_0 \, | \, {\bf x}_0) \, p_x^*({\bf x}_0)  \, d {\bf x}_0
\nonumber \\
& =  & \int_{{\mathbb R}^n} \Pi({\bf x}, t \, | \, {\bf x}_0, 0) \,
 p_x^*({\bf x}_0)  \, d {\bf x}_0  \, ,
\label{eq3_4}
\end{eqnarray}
where
\begin{equation}
\Pi({\bf x}, t \, | \, {\bf x}_0, 0) =
 \int_{{\mathbb R}^{n_h}} d{\bf q} \int_{{\mathbb R}^{n_h}}
p({\bf x},{\bf q}, t \, | \, {\bf x}_0, {\bf q}_0, 0) \,
p^*({\bf q}_0 \, | \, {\bf x}_0) \, d {\bf q}_0
\label{eq3_5}
\end{equation}
represents the equivalent of eq. (\ref{eq3_2}) in the
non-Markovian case.

\subsection{Non-equilibrium preparations}
\label{sec2.1}

The path-integral formulation of equilibrium conditions
expressed by eqs. (\ref{eq3_2}) or (\ref{eq3_4})
admits interesting  interpretations in terms of particle
trajectories and provides a way for extracting
valuable information about non-equilibrium
dynamics out of equilibrium trajectories.

\begin{figure}
\includegraphics[width=12cm]{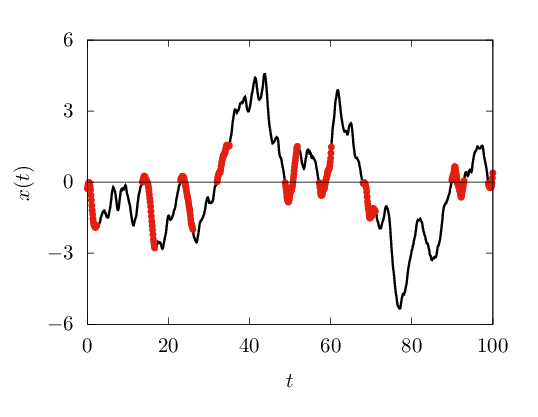}
\caption{Schematic representation of the subsampling procedure of
an equilibrium trajectory to obtain  non-equilibrium paths with
prescribed initial preparation, in the
present case starting from $x_0=0$. The non-equilibrium trajectories
starting from $x_0=0$ are marked with red ($\bullet$)-symbols.}
\label{Fig0}
\end{figure}

To begin with, consider the Markovian case. For Brownian
motion it corresponds to the Einstein-Langevin picture \cite{langevin,chandra}, where
the process ${\bf X}(t)$ is represented by the
particle position $X(t)$ and velocity $V(t)$,
\begin{equation}
{\bf X}(t)=
\left (
\begin{array}{cc}
X(t) \\
V(t)
\end{array}
\right )
\label{eq3_6}
\end{equation}
In this case, given an equilibrium trajectory expressed 
by the couple $(X(t),V(t))$, one can prepare the system
by considering a subsampling of the equilibrium trajectory, dissecting it into
a family of non-equilibrium trajectories originating
at the nominal rescaled time $t=0$ from one and the
same value of ${\bf x}_0={\bf X}(t=0)$, and
averaging over this family of subsampled trajectories.
In this way, the propagator $p({\bf x}, t \, | \, {\bf x}_0,0)$
is recovered. The pictorial representation of this
procedure is schematized in figure \ref{Fig0}.

In the non-Markovian case, provided the existence of a Markovian
embedding, the auxiliary variables ${\bf Q}(t)$ are not
directly available. While it is possible to infer some information about these hidden degrees of freedom from the past trajectory of the accessible degrees of freedom \cite{muruga_extracting_2026}, experimental conditioning of specific initial conditions for several auxiliary variables is yet to be demonstrated.  Accordingly, the
preparation of the system via a subsampling of an equilibrium
trajectory corresponds to a choice of a specific value
${\bf x}_0$ of ${\bf X}(t)$, at the nominal time $t=0$,
while the auxiliary variables ${\bf Q}(t=0)={\bf q}_0$
represent random variables distributed according
to the marginal equilibrium  distribution $p_{\bf q}^*({\bf q}_0)$.

The latter procedure provides a clear, experimentally-based, interpretation
of the variables ${\bf X}(t)$ and ${\bf Q}(t)$ in Brownian
motion theory: the ${\bf X}$-variables correspond to ``{\em the observables}''
of Brownian motion dynamics, i.e. particle position and velocity,
while the ${\bf Q}$-variables
play the role of ``{\em the hidden variables}'' of the theory,
as they are not amenable to a direct preparation in the subsampling
of equilibrium trajectories.
The decomposition
into ``observables'' and ``hidden variables'' is  operational
and well-posed, but it relies  on the observer. In this
meaning it constitutes a relational property  \cite{rovelli} of the system
Brownian-particle/observer that depends on the observer's
experimental apparatus and measurement resolution.
As addressed in Section \ref{sec3}, in Brownian motion hydrodynamics, the
hidden variables
correspond to the auxiliary memory variables
introduced in order
to account for the interactions of the particle with the
degrees of freedom of the solvent fluid and its inertial and relaxation
dynamics. This stems directly from the
modal expansion discussed in Section \ref{sec3}.

From eqs. (\ref{eq3_2}) or (\ref{eq3_4})  it is clear the
difference between the non-equilibrium states contemplated
throughout this article for a Brownian particle in a macroscopically
still fluid at constant temperature $T$ and confined by a harmonic optical
trap and the Non-Equilibrium Steady States (NESS) thoroughly analyzed
in the literature \cite{ness1,ness2,ness3,ness4}.
The former are dynamic states corresponding statistically
to the transition probability density $p({\bf x}, t \, | \, {\bf x}_0, 0)$
or $\Pi({\bf x}, t \, | \, {\bf x}_0, 0) $, in their  relaxation
towards  the equilibrium measure. Conversely, NESS are stationary states 
that arise in the presence of external forces that do not derive
from a potential. These forces lead to deviations from the equilibrium distribution at steady state and are associated with non-vanishing steady-state currents (fluxes).

\section{Brownian motion hydrodynamics}
\label{sec3}

Consider the case of  a micrometric spherical particle  of mass $m$ and radius
$R_p$ in a unconfined fluid medium at thermal equilibrium (constant
temperature $T$), subjected to a harmonic potential  stemming
from an optical trap. This corresponds to
the customary setting of highly resolved Brownian motion
experiments. The symmetry of spherical particle in an unconfined
medium allows us to consider a scalar description of particle
hydromechanics.

As extensively addressed elsewhere \cite{hydro1,hydro2,hydro3,hydro4,annurev}, 
the most general and comprehensive
formulation of Brownian hydromechanics in these
conditions  
involves the introduction of two hydromechanic
memory kernels, $h(t)$ and $k(t)$, so that
the equations of motion take the form
\begin{eqnarray}
\frac{d x(t)}{d t} & = & v(t)  \label{eq2_1} \\
m \, \frac{d v(t)}{d t} & = & - k_s \, x(t)-\int_0^t h(t-\tau) \, v(\tau) \, d \tau
- \int_0^t k(t-\tau) \, \left ( \frac{d v(\tau)}{d \tau} + v(0) \, \delta(\tau)
\right ) \, d \tau + R(t) \, ,
\nonumber
\end{eqnarray}
where $k_s$ is the spring constant of the trap and $R(t)$ the fluctuational
thermal-hydrodynamic force defined by the fluctuation-dissipation relations.
Brownian experiments in the last two decades have indeed shown the
reliability of the hydrodynamic approach \cite{hres1,hres2,hres3,hres4,btr}, 
that provides a physically-based expression for the kernels
in terms  either of the structural properties
of the particles (for non-spherical particles, the
kernels are matrix-valued \cite{kim,brenner}) and of the rheological
properties of the fluid (Newtonian, Maxwell fluids,  etc.).
The impulsive term entering eq. (\ref{eq2_1}) stems from the
inverse of the Laplace transform of the fluid-inertial
contribution in a Newtonian fluid defining the
Basset force \cite{kim}, and is further addressed in \cite{giona_nonergo}.
The second eq. (\ref{eq2_1}) can be compactly expressed by means of a single
global memory kernel $H(t)$,
\begin{equation}
m \, \frac{d v(t)}{d t}  =  - k_s \, x(t)-\int_0^t H(t-\tau) \, v(\tau) \, d \tau  + R(t) \, .
\label{eq2_1x}
\end{equation}

The two memory kernels entering eq. (\ref{eq2_1}), namely
the dissipative $h(t)$ and the fluid-inertial $k(t)$ kernels,
stem from the hydrodynamics of the solvent fluid and derive from
the assumptions regarding the  hydrodynamic regime and the rheological properties
of the fluid medium.
In a Newtonian fluid possessing viscosity $\mu$,  assuming the
Stokes regime for the solvent fluid, the fluid responds
instantaneously to external and boundary perturbations. Thus, $h(t) = \eta \, \delta(t)$,
so that the dissipative kernel is impulsive and is characterized exclusively by the
Stokesian friction factor $\eta=6\, \pi \, \mu \, R_p$. In this regime, the fluid-inertial kernel vanishes identically, $k(t)=0$.

Fluid-inertial effects should be properly taken into
account in the short-time Brownian dynamics \cite{hydro1,hydro2,annurev,btr}. 
This implies the consideration of the linear time-dependent Stokes regime for the solvent fluid. In the Newtonian incompressible case,
we still have $h(t) = \eta \, \delta(t)$, while
$k(t)= b/\sqrt{t}+m_a \, \delta(t)$ with $b=6 \, \sqrt{\pi \, \rho \, \mu} \, R_p^2$,
where $\rho$ is the fluid mass density and $m_a=2  \, \pi  \rho \, R_p^3/3$
\cite{kim}.
The importance of the fluid inertial contributions in short-time
Brownian dynamics has been confirmed and supported by several 
and different experimental analyses \cite{hres2,hres3,annurev,btr}.
 The fluid-inertial kernel
in this case  is made by two contributions: a continuous $1/\sqrt{t}$-kernel,
singular at $t=0$, corresponding to the Basset force, and the
impulsive contribution $m_a \, \delta(t)$ associated with the
added-mass $m_a$. Both these two singularities (localized at zero for the
Basset kernel,
and impulsive as regards the added-mass term) 
are mathematical artifacts of the assumption of infinite
propagation velocity of the hydrodynamic fields, and specifically
of the shear and compressible stresses \cite{gpp_pof}.
The first artifact, i.e. the singularity of $1/\sqrt{t}$ of the
Basset kernel is resolved by including viscoelastic effects that determine
a bounded propagation of the shear stresses.
All liquids do possess viscoelastic properties, but their
characteristic relaxation times can be so small
(order of 1 ps for water  \cite{waterrelax}) that they can be neglected in
practical microparticle transport at time scales equal to or  greater than
the particle momentum relaxation time $\tau_{p}=m/\eta$ (order
of $10^{-6}$-$10^{-7}$ seconds for a micrometric buoyantly neutral particle
in water). Regardless of the fact that the relaxation time may be small,
the intrinsic viscoelastic nature of any liquid 
reemerges in the
functional structure of the inertial kernel $k(t)$ as, owing
to the finite propagation of the shear stresses, the value
$k(0)$ at $t=0$ turns out to be bounded, resolving the singularity
of the Basset kernel  associated with the Newtonian rheological
assumption \cite{procopio_giona}.
For a Maxwell fluid characterized by a relaxation rate $\lambda$
(reciprocal of the relaxation time), we have $h(t)= \eta \, \lambda \, e^{-\lambda \,
t}$ and $k(t)=  b \, \sqrt{\lambda} e^{-\lambda t} \, I_0(\lambda  \, t/2)$ 
where $I_0(t)$ is the modified
Bessel function of the first kind and of zero order \cite{procopio_giona}.
Thus $k(0)= b \, \sqrt{\lambda}$.
The added-mass effect stems from the infinite velocity of propagation of sound
waves \cite{zwanzig}. If a finite  sound velocity is accounted for, 
then $m_a=0$ \cite{chow_hermans},
while the resulting kernel results in a slight modification of the
Basset  $1/\sqrt{t}$ that is immaterial as regards to the qualitative properties of Brownian dynamics.

Expressed in terms of the memory kernels $h(t)$ and $k(t)$,
and assuming that $h(t)=\eta \, \delta(t) + h_{\rm con}(t)$
where $h_{\rm con}(t)$ is the continuous non-impulsive part
of the dissipative response,  while $k(t)$ is
a continuous function of its argument, the fluctuation-dissipation
relations of the second-kind (adopting the terminology
introduced by Kubo \cite{kubo1,kubo2}) reads as
\begin{equation}
\langle R(t) \, R(0) \rangle_{\rm eq}=
\lim_{\tau \rightarrow \infty}\langle R(t+\tau) \, R(\tau) \rangle
= k_B \, T \, \left [ 2 \, \eta \, \delta(t)+ h_{\rm con}(t)
+ 2 \, k(0)\, \delta(t) + \frac{d k(t)}{d t} \right ],
\label{eq_k1}
\end{equation}
where  $k_B$ is the Boltzmann constant. From eq. (\ref{eq_k1})
the  mathematical necessity of a finite value for $k(0)=k(t=0)$
can be further appreciated in order to define the
thermal-hydrodynamic fluctuational force $R(t)$ \cite{gionacoccoprocopio}.
With eq. (\ref{eq_k1}) in mind,
a reliable approximation  for the global
kernel $H(t)$, obtaining
qualitatively and quantitatively accurate expressions
 for the short-time Brownian dynamics in a Newtonian
fluid including the influence of the Basset force, is  
\begin{equation}
H(t)= \eta \, \delta(t) + \frac{d k(t)}{d t} =
\eta \, \delta(t) - \frac{b}{2} t^{-3/2}.
\label{eq_k2}
\end{equation}
This expression will be applied in the next section, and
it is further discussed in Appendix \ref{app1}.
 
As extensively discussed in \cite{gpp1,gpp2,gpp3,gpp_pof} (see also
\cite{checkin,modal1,hanggi,giona_pre} for generic GLE), a convenient way for
approaching Brownian hydromechanics
both theoretically and computationally is to develop
the kernels $h(t)$ and $k(t)$ as a series of exponentially
decaying modes,
\begin{equation}
h(t)= \sum_{i=1}^{N_d} h_i \, \lambda_i \, e^{-\lambda_i \, t}
+ \eta_0 \, \delta(t) \,, \quad k(t)=\sum_{\alpha=1}^{N_i} \gamma_\alpha
\, e^{-\mu_\alpha \, t},
\label{eq2_2}
\end{equation}
where $h_i\,, \lambda_i >0$, $i=1,\dots N_d$, $\gamma_\alpha \, , \mu_\alpha >0$,
$\alpha=1,\dots,N_d$. As discussed in \cite{giona_nonergo,gpp2}, and further explicited in
Section \ref{sec4}, it is possible to recover a power-law scaling
of the kernels via a relatively small number of modes: on the order of a dozen.
The representation eq. (\ref{eq2_2}) is a
 classical Prony series expansion, widely used in rheological
analysis \cite{macosko}, extended also to the fluid inertial kernel.
It proves its major advantage when the  explicit expression
of the fluctuational force $R(t)$ should be obtained in terms
of elementary stochastic processes, as addressed in \cite{gpp1} and succinctly 
reviewed below.

In a fluid at thermal equilibrium the statistical
properties of the fluctuation force stem from
Kubo fluctuation-dissipation theory. The  modal expansion of the kernel
eq. (\ref{eq2_2}) via decaying exponential modes provides
a simple and effective setting to formulate Brownian motion
dynamics consistent with Kubo fluctuation-dissipation
theory. This is thoroughly explained in \cite{gpp1} and therefore it
is unnecessary to reiterate all the details.
Essentially, it is possible to introduce
a system of auxiliary modes $\{ \theta_i(t)\}_{i=1}^{N_d}$, 
and $\{z_\alpha(t) \}_{\alpha=1}^{N_i}$,  each of which characterized
by its proper relaxation rate, $\lambda_i$, $i=1,\dots,N_d$,
$\mu_\alpha$, $\alpha=1,\dots,N_i$, respectively,
such that the fluctuational thermal/hydrodynamic force $R(t)$
can be represented via the introduction
of impulsive forcings acting either on the particle velocity $v(t)$
itself, or on the auxiliary memory variables $\{ \theta_i(t) \}_{i=1}^{N_d}$
and $\{ z_\alpha(t) \}_{\alpha=1}^{N_i}$.
The resulting general formulation of trapped Brownian motion dynamics
in the presence of the kernels defined by eq. (\ref{eq2_2})
is given by the system of Langevin equations
\begin{eqnarray}
\frac{d x(t)}{d t} & =  &v(t) \nonumber \\
m \, \frac{d v(t)}{d t}  & = & - k_s \, x(t) - \sum_{i=1}^{N_d} h_i \, \lambda_i
\, \theta_i(t) - \eta_0 \, v(t) - \left ( \sum_{\alpha=1}^{N_i}
\gamma_\alpha \right ) \, v(t)+
\sum_{\alpha=1}^{N_i} \mu_\alpha \, \gamma_\alpha \, z_\alpha(t)
\nonumber \\
&+ & \sqrt{2} \, a \, \xi(t) + \sqrt{2} \, \sum_{\alpha=1}^{N_i} d_\alpha
\,\xi_\alpha^{(i)}(t) \label{eq2_3} \\
\frac{d \theta_i(t)}{d t} & = & -\lambda_i \, \theta_i(t) + v(t) +
\sqrt{2} \, b_i \, \xi_i^{(d)}(t) \, , \quad i=1,\dots,N_d
\nonumber \\
\frac{d z_\alpha(t)}{d t} & =  & -\mu_\alpha \, z_\alpha(t) + v(t)
+ \sqrt{2} \, c_\alpha \, \xi_\alpha^{(i)}(t) \, , \quad \alpha=1,\dots, N_d
\nonumber 
\end{eqnarray}
where $\xi(t)$, $\xi_i^{(d)}(t)$ and $\xi_\alpha^{(i)}(t)$ are
distributional derivatives of independent Wiener processes,
\begin{eqnarray}
\langle \xi(t) \, \xi(t^\prime) \rangle  & = & \delta(t-t^\prime) \, ,
\quad \langle \xi(t) \, \xi_i^{(d)}(t^\prime) \rangle =
\langle \xi(t) \, \xi_\alpha^{(i)}(t^\prime) \rangle \nonumber \\
\langle \xi_i^{(d)}(t) \, \xi_\alpha^{(i)}(t^\prime) \rangle  & = &  0 
\label{eq2_4} \\
\langle \xi_i^{(d)}(t) \,\xi_j^{(d)}(t^\prime)  \rangle  & = &
\delta_{ij} \, \delta(t-t^\prime) \, , \quad
\langle \xi_\alpha^{(i)}(t) \,\xi_\beta^{(i)}(t^\prime)  \rangle =
\delta_{\alpha \beta} \, \delta(t-t^\prime) \nonumber
\end{eqnarray}
for any $t,t^\prime >0$, and for any $i,j=1,\dots,N_d$,
$\alpha,\beta=1, \dots, N_i$. The coefficients $a$, $b_i$, $i=1,\dots, N_d$,
$c_\alpha, \, d_\alpha$, $\alpha=1,\dots,N_i$, entering eqs. (\ref{eq2_3})
are uniquely defined by the Kubo fluctuation-dissipation relations.
The details of this calculation can be found in \cite{gpp1} and are
explicitly reviewed in the remainder whenever  needed.

The choice of Wiener processes to define the
elementary forcings $\{ \xi_i^{(d)}(t) \}_{i=1}^{N_d}$,
$\{\xi_\alpha^{(i)}(t) \}_{\alpha=1}^{N_i}$ stems from
the fact that we are considering Brownian motion in liquids
for which the  fluctuational force $R(t)$ reasonably admits
a continuous support in time. We mention this detail as $R(t)$ can
be  expressed in terms of elementary forcings  corresponding to
the distributional derivatives of compound Markov and semi-Markov 
counting processes \cite{gpp2} that are impulsively
different from zero on a numerable set of time instants. This representation is more suitable
for Brownian motion in extremely diluted gaseous media (medium
or high vacuum), and  the  short-time dynamics in these
conditions is different from the liquid case treated in the remainder. 

Eq. (\ref{eq2_3}) corresponds to the formulation developed in \cite{gpp1} 
and referred to as  a fluctuational pattern,  consistent with
classical fluctuation-dissipation theory that, as follows from
eq. (\ref{eq2_4}), is excited by elementary $\delta$-correlated
stochastic forcing entering the evolution equations
for $v(t)$, $\{\theta_i(t)\}_{i=1}^{N_d}$ and $\{ z_\alpha(t) \}_{\alpha=1}^{N_i}$ in a way to fulfill Kubo fluctuation-dissipation relations
of the first and second kind \cite{kubo1}.

A further generalization stems from the consistency principle
of requiring bounded propagation of all the hydrodynamics quantities,
including the stochastic forcings, in line with  the same requirement
invoked for  
 shear and compression stresses.
This  principle leads to the concept of correlated fluctuational
patterns, in which the impulsive stochastic processes
$\xi(t)$, $\xi_i^{(d)}(t)$, $i=1,\dots,N_d$,  $\xi_\alpha^{(i)}(t)$,
$\alpha=1,\dots,N_i$ are substituted by normalized correlated
processes possessing a finite non-zero correlation time
(for the details of the normalization see \cite{gpp3}).
The salient feature of the correlated fluctuational patterns
is that they still fulfill the global fluctuation-dissipation
relation (Stokes-Einstein relation), connecting particle diffusivity,
i.e. the integral of the velocity autocorrelation function,
to the global friction coefficient.

\section{Non-equilibrium short-time dynamics}
\label{sec4}

The preparation of initial conditions via the subsampling of equilibrium
trajectories permits the extension of analysis to non-equilibrium
dynamics, providing a way to obtain detailed information
on fluid-particle interactions at microscale directly from
the analysis of the lower-order moments of the particle position.

The analysis of the lower-order moments is aimed at
 obtaining  a qualitative and quantitative discrimination of 
the hydrodynamic regimes controlling Brownian dynamics.
To begin with, the analysis is carried out starting
from the GLE formalism using the free inertial process approach, 
and then developed via
modal expansion, to  support some technical issues,
and to provide a general characterization
of hydrodynamic regimes and their short-time signatures.
To focus on the physical aspects, we discuss
in the main text the principal results, leaving to
Appendix \ref{app2} a thorough  description  of the formal details.

To avoid confusion, we use throughout the article the symbol ``$\sim$''
to indicate scaling, so that $f(t) \sim t^\alpha$ for $t \rightarrow 0$
indicates that there exists a constant $c \neq 0$ such tha
$\lim_{t \rightarrow 0} f(t)/t^\alpha=c$. Conversely, the
symbol ``$\simeq$'', i.e. $f(t) \simeq  c \, t^\alpha$  for $t \rightarrow 0$
indicates that
this expression is correct to the leading order i.e. that there exists an $\varepsilon>0$ such that $f(t)= c \,  t^\alpha + O(t^{\alpha+\varepsilon})$.

\subsection{Free inertial process approach}
\label{sec4_1}

It has been shown  in \cite{duplat,wm,btr}
that the short-time  super-ballistic scaling of the mean square
displacement  is 
dependent on the color of the thermal force interacting with the particle. 
In \cite{masoliver_harmonic_1993, btr}, it has been  argued that the leading order, zero initial 
velocity, mean square displacement
 can be derived from a free inertial process with corresponding 
stochastic differential equation
\begin{equation}
    \label{free_inertial}
    m \, \frac{d^2 x(t)}{d t^2} = R(t)
\end{equation}
The justification of this result is  that the other terms should be 
unimportant at small times since the velocity is initially vanishing and 
it has been shown in \cite{btr} that it provides the correct results to 
first order for both the white noise and hydrodynamic models (i.e.
including fluid inertia). The same argument is presented in \cite{masoliver_harmonic_1993, wm} in the case of finite memory kernels in the short-time limit and the authors study such processes further in \cite{masoliver_free_1995}.
Below, the validity of eq. (\ref{free_inertial})
is further justified in greater detail
addressing some subtleties that arise in the hydrodynamic case. 
Consider a GLE with some general kernel $H(t)$, i.e. eq. 
(\ref{eq2_1x}). Without loss of generality assume $k_s=0$,
simply because the effect of the harmonic
trap is negligible starting from 
an initial preparation of the system with $x_0=v_0=0$.
We leave $H(t)$ as a general physically realizable memory kernel.
From eq. (\ref{eq2_1x}) with $k_s=0$, indicating
with $\widehat{v}(s)={\mathcal L}[v(t)]$ the Laplace transform
of $v(t)$ (and similarly for the other quantities),
we have
\begin{equation}
    \label{Laplace_domain}
    \widehat{v}(s) = \frac{\widehat{R}(s)}{ms}\frac{1}{1+\frac{\widehat{H}(s)}{ms}}
\end{equation}
The short-time dynamics in the time domain are set by the large $s$ values of 
$\widehat{v}(s)$ in the Laplace domain. As a consequence of the theoretical analysis developed in \cite{fluids9110260}, if we ignore any added mass effects of the fluid inertial coupling, 
\begin{equation}
    \lim_{s \rightarrow \infty} \frac{\widehat{H}(s)}{s} = 0.
\label{eq_jc1}
\end{equation}
The added mass component of the coupling scales like $s$, and therefore does not in principle obey the above relation. However, because its contribution is indistinguishable from the inertia of the particle, we can always incorporate its effect into a modified mass value of the particle such that the following analysis holds. Thus, for sufficiently large $s$ (or sufficiently small $t$), we can make 
$\widehat{H}(s)/{m \, s}$ a small parameter, and we can expand eq. (\ref{Laplace_domain}) in powers of this small 
parameter. This yields 
\begin{equation}
    \widehat{v}(s) = \frac{\widehat{R}(s)}{m \, s} 
\left [ 1 - \frac{\widehat{H}(s)}{m \, s} + 
\left ( \frac{\widehat{H}(s)}{m \, s} \right )^2 - 
\left (\frac{\widehat{H}(s)}{m \, s} \right )^3+... \right ]
\label{eq_jc2}
\end{equation}
The first term in this expansion corresponds exactly with the Laplace 
domain form of eq. (\ref{free_inertial}). Thus, there exists some short time period during which the 
dynamics can be described by eq. (\ref{free_inertial}). 
This provides a powerful tool to understand dynamics for a variety of memory 
kernels
including the effect of the hydrodynamic Basset kernel.
In the latter case, enforcing the  Kubo fluctuation-dissipation
relation of the second kind and considering 
 eq. (\ref{eq_k2}) for $H(t)$,
the thermal-hydrodynamic  force $R(t)$ possesses a 
correlation function at equilibrium defined by 
\begin{equation}
    \langle R(t_1) \, R(t_2)\rangle = 2 \, \eta k_B  \, T 
\left (\delta(t_1 - t_2) - \frac{1}{4}\sqrt{\frac{\tau_\nu}{\pi}} 
\, |t_1 - t_2|^{-3/2} \right )
\label{eq_jc3}
\end{equation}
where $\tau_\nu=R_p^2/\nu$, $\nu=\mu/\rho$ being the
kinematic viscosity of the fluid, is the characteristic
time of the viscous propagation.
Solving eq. (\ref{free_inertial})
 for the above correlation function, one obtains 
\begin{equation}
    \langle x^2(t)\rangle  =\frac{2}{3}\frac{k_B \, T}{ m \,
 \tau_p}  \left (\frac{12}{5}\sqrt{\frac{\tau_\nu}{\pi}}t^{5/2} +  t^3
\right )
\label{eq_jc4}
\end{equation}
where $\tau_p=m/\eta$. Note that to solve this differential equation,
 we rely on an analytic continuation of the Laplace transform such that $\mathcal{L}\left [ d (t^{-1/2})/dt\right] = 2\sqrt{\pi s}$.
If instead we consider the velocity autocorrelation
function starting from $v_0=0$, expanded for small $t$, and integrate
 it to obtain $\langle x^2(t)\rangle$, 
one recovers the result obtained in  \cite{btr}, namely
\begin{equation}
     \langle x^2(t)\rangle  \simeq\frac{2}{3}\frac{k_B \, T}{m \,  \tau_p} 
\left (\frac{12}{5} \sqrt{\frac{\tau_\nu}{\pi}}  \, t^{5/2} + c \,
 t^3 \right) \, , \quad
c = 1 - \left (1 + \frac{8}{3\pi} \right ) \, \frac{\tau_\nu}{\tau_p}
 \, . \label{two_term_expansion} 
\end{equation}
The difference  between eqs. (\ref{eq_jc4}) and (\ref{two_term_expansion})
can be accounted for by including the next highest order term in the 
expansion of the Green's function in the Laplace domain. Keeping the two 
lowest order terms in eq. (\ref{eq_jc2}) yields
\begin{equation}
\widehat{v}(s) \, \widehat{v}(s^\prime) \simeq \frac{\widehat{R}(s) \, 
\widehat{R}(s')}{m^2 \, s \, s^\prime} \left
(1- \frac{\widehat{H}(s)}{m \, s} - \frac{\widehat{H}(s^\prime)}{m\, s^\prime}
\right ) \, .
\label{eq_jc5}
\end{equation}
For a free particle in a dense fluid initiated with vanishing velocity,
\begin{equation}
    \widehat{v}(s) = \frac{R(s)}{m \, s +\eta \sqrt{\tau_\nu \, s}+\eta}
\, .
    \label{v_eq}
\end{equation}
Thus, keeping to the next lowest order yields
\begin{equation}
    \widehat{v}(s) \simeq \frac{\widehat{R}(s)}{m \, s}
\left (1-\frac{\eta \sqrt{\tau_\nu \,  s}}{m \, s} \right ) \, .
\label{eq_jc6}
\end{equation}
Solving for the velocity autocorrelation function,
with $\widehat{v}(s)$ given by eq. (\ref{eq_jc6}) one obtains 
\begin{equation}
    \langle \widehat{v}(s) \, \widehat{v}(s^\prime) \rangle 
\simeq \frac{\langle \tilde{R}(s)\tilde{R}(s') \rangle}{m^2 \, s 
\,s^\prime} \, \left  (1 - \frac{\eta \sqrt{\tau_\nu \,  s}}{m\, s}
\right ) \left (1 -\frac{\eta \sqrt{\tau_\nu \, s^\prime}}{m \, 
s^\prime} \right ) \, .
\label{eq_jc7}
\end{equation}
Ignoring terms of order $1/s$ and the purely inertial term and using the expression for $\langle \widehat{R}(s) \, \widehat{R}(s^\prime) \rangle_{\rm eq}$ 
found in \cite{fox_gaussian_1978} gives
\begin{equation}
  \langle \widehat{v}(s) \, \widehat{v}(s^\prime)\rangle^{\rm (1)} 
= \frac{k_B  \, T \eta^2 \tau_\nu}{m^3} \left ( - \frac{s}{s^2  \, s^\prime \,
(s+s^\prime)} - \frac{s^\prime}{s \, s'^2(s+s^\prime)} - 
\frac{\sqrt{s  \, s^\prime}}{s^2 \, (s^\prime)^2} \right ) \, ,
\label{eq_jc8}
\end{equation}
where $\langle \cdot \rangle^{(1)}$ indicates the first order correction 
of this correlation to the free inertial theory. Inverting this expression to the time domain yields
\begin{equation}
    \langle v(t_1) \, v(t_2) \rangle^{\rm (1)}
 = -\frac{k_B  \, T }{m^3 \, \tau_p}\frac{\tau_\nu}{\tau_p} \, \left (2 \,
\mbox{min}(t_1, t_2) + \frac{4}{\pi}\sqrt{t_1 t_2} \right ) \, .
\label{eq_jc9}
\end{equation}
Integrating this expression twice yields another term that goes like $t^3$ and gives the correct value from theory for $c$ entering eq. (\ref{two_term_expansion}). Thus, the tension between the two results is resolved in agreement with the experimental results in \cite{btr}. Specifically, dynamics are driven both by dissipative delta-correlated Stokesian and the anti-persistent inertial thermal fluctuations. The second order inertial effect is of the same order as the first order dissipative effect and therefore this second order effect modifies the coefficient of the $t^3$ term obtained from the free inertial process.

Next, consider any GLE of the form of eq. (\ref{eq2_1x})
 such that $H(t)$ is finite and smooth as $t \to 0$. 
This should be true for any physically realizable memory kernel. 
Then, for some time interval around $t=0$, we can write $H(t)$ in Taylor 
series 
\begin{equation}
    H(t) = c_0 +c_1 \, t+c_2 \, t^2 + O(t^3) \, .
\label{eq_jc10}
\end{equation}
Then, over the time interval where the particle's motion can be 
approximated by the free inertial process eq. (\ref{free_inertial}),
\begin{equation}
    \langle x^2(t)\rangle = \frac{1}{m^2}\int_0^t dt_1 
\int_0^tdt_2 \int_0^{t_1}dt_1^\prime \int_0^{t_2} dt_2^\prime
 \langle R(t_1^\prime)R(t_2^\prime) \rangle  \, .
\label{eq_jc11}
\end{equation}
By the fluctuation dissipation theorem,
\begin{equation}
\langle R(t_1^\prime)R(t_2^\prime)\rangle = k_B \, T \,
H(|t_1^\prime - t_2^\prime|) =  k_B \, T \, \left [H(0) + H^\prime(0)\, 
|t_1^\prime -t_2^\prime| + O(|t_1^\prime - t_2^\prime|^2) \right ] \, ,
\label{eq_jc12}
\end{equation}
where $H^\prime(0)=dH(t)/dt|_{t=0}$.
Therefore,  for small enough $t$,
\begin{equation}
    \langle x^2(t)\rangle \simeq  \frac{1}{m^2} \int_0^t dt_1
\int_0^tdt_2 \int_0^{t_1} dt_1^\prime \int_0^{t_2} H(0) \, dt_2^\prime 
\simeq \frac{H(0) \, t^4}{4 \, m^2} \, .
\label{eq_jc13}
\end{equation}
This result for colored $R(t)$ not possessing a white component
has been derived  in \cite{wm}, using similar arguments, and by
Kang et al. \cite{quantum} for diffusion of a quantum
particle in the presence of a correlated noise potential.
The latter result is not surprising, owing to the
Ehrenfest theorem for the expected values of quantum motion
and its relation to classical trajectories.

In a similar way, if $H(t) \sim t^{-\beta}$, $\beta \in [0,2)$,
in the neighbourhood of $t=0$, then  $\langle x^2(t)\rangle \sim
t^{4-\beta}$. Note this scaling agrees with the analysis presented \cite{sandev_langevin_2014, tateishi_different_2012} for the short-time growth in fluctuations around the mean trajectory for fractional Langevin equations. Thus, by expansion in the Laplace domain, we formalize the arguments from  \cite{wm, btr} and demonstrate how to derive higher-order corrections to the free inertial 
approach from the Laplace domain expansion. We further demonstrate the value of this approach by considering trajectories conditioned on thermal force values.

\subsection{Scaling for conditioned thermal force}
\label{app3}
Let us demonstrate both the efficacy of the free inertial process approach as well as the scaling properties of Z-preparation states. Here, we extend the use cases of the free-inertial systems to include those in which the thermal force is also conditioned on a specific initial value. Of particular interest is the case where the thermal force is also conditioned to begin close to 0. Assuming that our thermal force $R(t)$ is a Gaussian random process, we can still utilize the free inertial approach to solve the future dynamics but with a modified thermal force with statistics defined by
\begin{equation}
    \label{cond_therm_force}
    \langle R(t_1)R(t_2) | R(0)=0 \rangle = \langle R(t_1)R(t_2)\rangle - \frac{\langle R(t_1)R(0)\rangle \langle R(t_2)R(0)\rangle}{\langle R(0) R(0)\rangle} 
\, .
\end{equation}
In the case of a memory kernel $H(t)$ that is well behaved close to the origin such that we can Taylor expand, this gives
\begin{multline}
    \langle R(t_1)R(t_2) | R(0)=0 \rangle =  k_B \, T \,  [H(0) + H^\prime(0)\, 
|t_1 -t_2| + O(|t_1 - t_2|^2)  ]\\ -k_B \, T \,\frac{([H(0) + H^\prime(0)\, 
t_1 + O(t_1^2)][H(0) + H^\prime(0)t_2 + O(t_2^2)]}{H(0)} \, .
\end{multline}
Solving to first order then yields
\begin{equation}
   \langle R(t_1)R(t_2) | R(0)=0 \rangle  \simeq k_B \, T \,H^\prime(0)[|t_1-t_2| - (t_1+t_2)] \, .
\end{equation}
Utilizing the free inertial process approach implies that
\begin{equation}
    \label{t5_scaling}
    \langle x^2(t)\rangle \simeq -k_B \, T \,H^\prime(0)\frac{t^5}{10} \, .
\end{equation}
Note for memory kernels of interest we expect $H'(0)<0$ so that we achieve a positive value for $\langle x^2(t)\rangle$.
As an example, we consider the case of a memory kernel consisting of a single exponential 
mode, $H(t)=Ae^{-a t}$. Descriptions for the evolution of such systems in equilibrium are well 
developed \cite{wm}. Under the formalism of eq. (\ref{cond_therm_force}), we find for an exponentially correlated noise undergoing a free inertial process
\begin{eqnarray}
    \langle x^2(t)|v(0) &= & 0, R(0)=0 \rangle \approx\frac{A}{m^2} 
\left (\frac{2t}{a^3}(1-e^{-at})-\frac{t^2}{a^2}+\frac{2}{3}\frac{t^3}{a} +\frac{2 - 2 e^{-at}- 2 a t}{a^4}
\right )  \nonumber  \\ &-  & \frac{A}{m^2 a^4} \left (e^{-at} -1+at \right )^2
\label{eqja1}
\end{eqnarray}
\begin{figure}
\includegraphics[width=12cm]{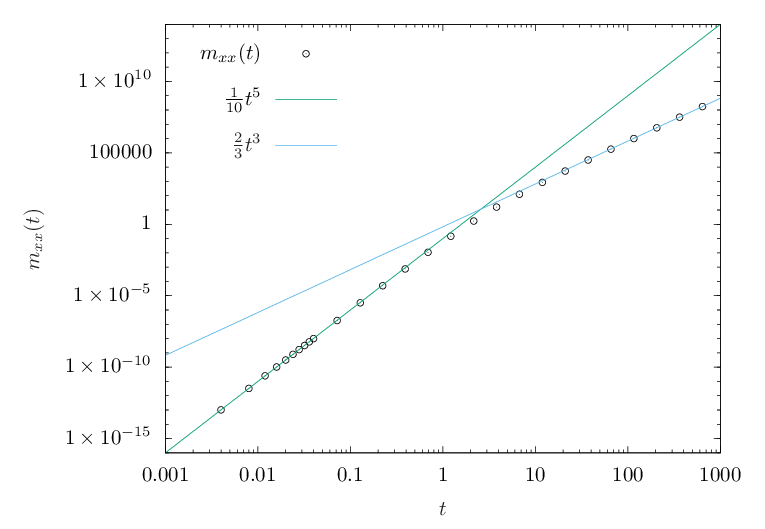}
\caption{$m_{xx}(t)=  \langle x^2(t)|v(0) =  0, R(0)=0 \rangle$ vs t expressed
by eq. (\ref{eqja1}) for  zero velocity and zero initial thermal force conditions
for a GLE governed by a single exponential mode.}
\label{exp_mode_0_R} 
\end{figure}
We plot this expression with $A=a=k_BT=1$ in Fig. \ref{exp_mode_0_R} along with two scalings. 
Below the correlation time of the memory kernel, we see the $t^5$ scaling predicted in eq. (\ref{t5_scaling}). 
Above the correlation time of the memory kernel, we see a $t^3$ scaling equivalent to the short-time predictions we expect for a system driven by white noise. \par
One can also ask what would happen if we condition on the value of the thermal force corresponding to the rational memory kernels that we see in the case of fractional Langevin equations. 
The singularity of these models at the origin generates difficulty for 
direct application of eq. (\ref{cond_therm_force}). However, as discussed throughout the text, any physically 
realizable memory kernel should be regularized at short enough time scales. 
As an explicit example of these sorts of models, we consider the Markovian embedding of the hydrodynamic 
Basset kernel developed in paragraph \ref{sec4_2}. Then, $\langle R(0)R(0) \rangle = 1/\sqrt{t_c}$ and 
\begin{equation}
    \langle R(t_1)R(t_2)|R(0)=0\rangle = \langle R(t_1)R(t_2)\rangle -\sqrt{t_c}\langle R(t_1)R(0)\rangle\langle R(t_2)R(0)\rangle  \, .
\end{equation}
Note that the experimentally accessible time scales are well above $t_c$, thus for observable time series, 
we expect $\langle R(t)R(0)\rangle = 1/\sqrt{t}$. However, since again we expect these timescales to be several orders of magnitude greater than $t_c$, we expect $\sqrt{t_c}\langle R(t_1)R(0)\rangle << 1$. Thus, as long as we are operating on time scales where the memory kernel is well approximated by the fractional scaling, we expect the conditioned thermal force to appear comparable to the equilibrium form.

\subsection{Modal analysis}
\label{sec4_2}

While the leading-order expansion of the
overall kernel $H(t)$ (which coincides with a Taylor-series
expansion if $H(t)$ is smooth in the neighbourhood of $t=0$)
provides a simple and powerful tool to assess
the short-time scaling of particle displacement statistics
from prepared non-equilibrium conditions ($x_0=v_0=0$),
some further refinements are required in order to
establish the theory unambiguously and connect it to 
the hydrodynamic regimes experienced by the Brownian particle.

As discussed in Appendix \ref{app1}, the structure of the overall kernel
$H(t)$ defined by eq. (\ref{eq_k2}) for a Newtonian
fluid in the presence of inertial effects (Basset force)
is not fully consistent, as the $t^{-3/2}$ contribution
provides unphysical results with regards to dissipation.
This inconsistency involves the long-term behavior in the time domain representation and thus does not appear in previous analysis via analytic continuation in the Laplace transform.

To clear the fog of doubt related to this issue, it is
convenient to enforce a modal representation of 
particle dynamics \cite{gpp1}, expanding $h(t)$ and/or $k(t)$
in Prony series eq. (\ref{eq2_2}).  
In order to keep the article focused on the
physical issues, we present in this paragraph the main results associated
with a singular kernel, referring to Appendix \ref{app2}
for a thorough presentation and description of these
models. We consider a non-dimensional formulation
of particle dynamics where time is rescaled with respect to
the momentum relaxation time, and velocities to
the characteristic thermal velocity $\sqrt{k_B \, T/m}$.
Henceforth, the  moment $m_{xx}(t)$ corresponds to the 
second-order moment of particle position $x$, conditional
to $x_0=v_0=0$.

\begin{figure}
\includegraphics[width=12cm]{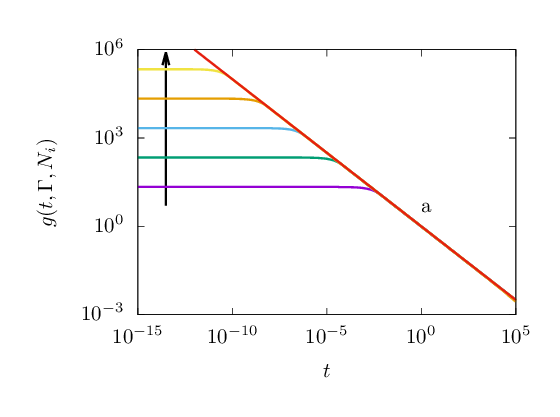}
\caption{$g(t,\Gamma,N_i)$  vs $t$. The arrow indicates
increasing values of $\Gamma=10^3,\,10^6,\,10^9,\,10^{11}$.
$N_i=20$ with the exception of $\Gamma=10^{11}$ where $N_i=30$.
Line (a) represents the function, $g(t,\Gamma,N_i)=1/\sqrt{t}$.}
\label{Fig2}
\end{figure}

The realization of power-law kernels can be obtained by
considering the family of functions $g_\xi(t,\Gamma)$
\begin{equation}
g_\xi(t,\Gamma)=  \sum_{k=1}^{\infty} \left ( \frac{\Gamma}{a^{k}}
\right )^\xi
e^{-t \, \Gamma/a^k}
\label{eqi_3}
\end{equation}
where $a>0$, $\Gamma>0$ and $\xi \in (0,1)$.
Set $a=6$,  and consider the case $g(t,\Gamma)=g_{1/2}(t,\Gamma)$.
Indicate with $g(t,\Gamma,N_i)$ the truncation of the
series eq. (\ref{eqi_3}) to $N_i$ modes.
The function $g(t;\Gamma)$, for $a=6$ behaves as 
\begin{equation}
g(t,\Gamma)= \left \{
\begin{array}{ccc}
\frac{\sqrt{\Gamma}}{\sqrt{6}-1} & \;\;\;\; \mbox{for  } t \ll t_c \\
 & \\
\frac{1}{\sqrt{t}}+O(t^{-3/2}) & \;\;\; \mbox{for  } t \gg t_c
\end{array}
\right .
\label{eqi_4}
\end{equation}
where $t_c$ is the crossover time 
\begin{equation}
t_c= \frac{(\sqrt{6}-1)^2}{\Gamma} ,
\label{eqi_5}
\end{equation}
as shown in figure \ref{Fig2}.

Therefore, for any $\Gamma$,  $g(t,\Gamma)$ possesses a $1/\sqrt{t}$ scaling
like the Basset force for $t>t_c$, where $t_c$ is order of  $0.1/\Gamma$.
Observe that in the present non-dimensional formulation $t=1$ corresponds
to the momentum relaxation time.

Moreover, for finite  $\Gamma$'s, the
limit for $t \rightarrow 0$ exists and is bounded as in any physical model
accounting for stress relaxation.
With $N_i \sim 30$, it is possible to recover these
behaviors over the time scales of physical interest using a relatively
small number of modes
 (see figure \ref{Fig2}).

\begin{figure}
\includegraphics[width=12cm]{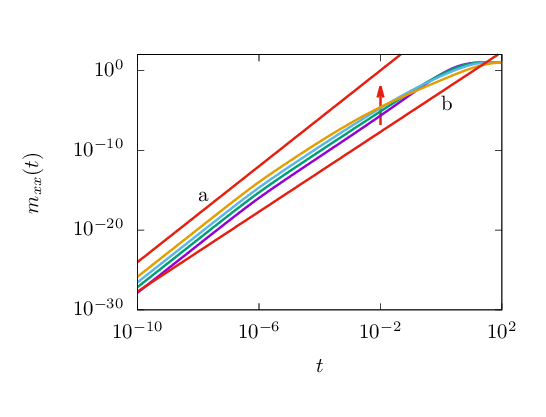}
\caption{$m_{xx}(t)$ vs $t$ for the Stokes-inertial model,
eq. (\ref{eqadd1}), with $\kappa_s=0.1$, and $\Gamma=10^7$
for different value of $b=0.1,\, 0.5,\, 2,\, 10$.
The arrow indicates increasing values of $b$. Line (a)
represents the initial scaling $m_{xx}(t) \sim t^3$,
line (b) the intermediate scaling $m_{xx}(t)\sim t^{5/2}$.}
\label{Fig3}
\end{figure}

\begin{figure}
\includegraphics[width=12cm]{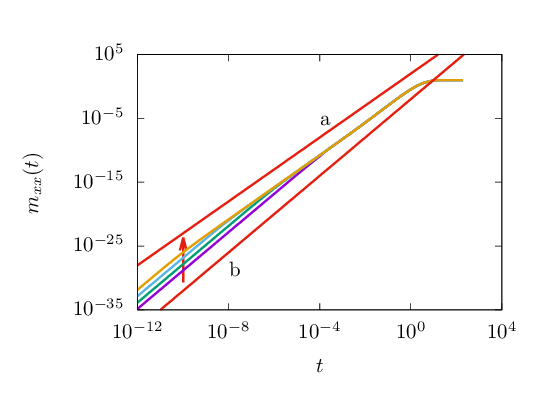}
\caption{$m_{xx}(t)$ vs $t$ for the Stokes-inertial model,
eq. (\ref{eqadd1}), with $\kappa_s=0.1$, $b=0.1$, for different
values of $\Gamma=10^5,\,10^7,\,10^9,\,10^{11}$.
The arrow indicates increasing values of $\Gamma$.
Line (a) represents the scaling $m_{xx}(t) \sim t^{5/2}$,
line (b) $m_{xx}(t) \sim t^3$.}
\label{Fig4}
\end{figure}

Consider the case of a Newtonian
fluid including the effect of the Basset force.
In this case, the nondimensional
kernels can be expressed as $h(t)=\delta(t)$, $k(t)=b \, g(t,\Gamma, N_i)$
for $\Gamma$ and $N_i$ sufficiently high. We use $N_i=30$, and
$\Gamma$ as a parameter.
The prefactor $b$ stems from hydrodynamics: for a spherical
particle  it is related
to the fluid ($\rho$) and particle  ($\rho_p$) densities
by  the relation $b=\sqrt{9 \, \rho/2 \, \pi \, \rho_p}$.
Its values may range from 0.1 to 3 for typical fluid and particle combinations.
The nondimensional modal representation of particle dynamics reads
\begin{eqnarray}
\frac{d x(t)}{d t} & = & v(t) \nonumber \\
\frac{ d v(t)}{d t} & = & - v(t) - \kappa_s \, x(t)- G \, v(t) +
\sum_{\alpha=1}^{N_i} \gamma_\alpha \, \mu_\alpha \, z_\alpha(t)
+ \sqrt{2} \,  \sum_{\alpha=1}^{N_i} d_\alpha \, \xi_\alpha(t) + \sqrt{2} \, \xi(t) \label{eqadd1} \\
\frac{d z_\alpha(t)}{d t} & = & - \mu_\alpha \, z_\alpha(t) + v(t) + \sqrt{2}
\, c_\alpha\, \xi_\alpha(t) \nonumber
\end{eqnarray}
where  $\xi(t)$ and $\xi_\alpha(t)$ and $\alpha=1,\dots,N_i$
are distributional derivatives of independent Wiener processes,
$G=\sum_{\alpha=1}^{N_i} \gamma_\alpha$,
$d_\alpha=\sqrt{\gamma_\alpha}$, $c_\alpha=-1/\sqrt{\gamma_\alpha}$, 
 $\alpha=1,\dots,N_i$.
The results are summarized in figures \ref{Fig3}-\ref{Fig4}.
It can be observed that for any finite value of $\Gamma$ the
initial scaling is always $m_{xx}(t) \sim t^3$,  followed
by are more or less pronounced intermediate scaling
 $m_{xx}(t) \sim t^{5/2}$. However,
as $\Gamma$ increases (figure \ref{Fig4}) the
scaling $m_{xx}(t) \sim t^{5/2}$ becomes more evident and
pronounced starting from early times. In the limit
for $\Gamma \rightarrow \infty$, corresponding to
the Basset kernel singular at $t=0$, the early scaling
is purely $m_{xx}(t) \sim t^{5/2}$.

This result fully support the analysis developed
in the previous paragraph based on the free inertial process approach.

Next, turn attention to the dissipative case possessing power-law
memory dynamics and neglecting the effects of fluid inertia,
i.e.  $h(t)= g_\omega(t,\Gamma,N_d)= \sum_{i=1}^{N_d} h_i \, e^{-\lambda_i \, t}$,
$k(t)=0$. Without loss of generality set $\kappa_s=0$ where
$\kappa_s$ is the nondimensional spring constant. The modal representation
of particle dynamics attains in this case the form
\begin{eqnarray}
\frac{d x(t)}{d t } & = & v(t) \nonumber \\
\frac{d v(t)}{d t}  & =  & - \sum_{i=1}^{N_d} h_i \, \theta_i(t)
\label{eqadd2} \\
\frac{d \theta_i(t)}{d t} &= & -\lambda_i \, \theta_i(t)+ v(t)
+ \sqrt{2} \, b_i \, \xi_i(t)
\nonumber
\end{eqnarray}
where $\xi_i(t)=d w_i(t)/d t$, $i=1,\dots,N_i$ are distributional
derivatives of independent Wiener processes $w_i(t)$,
and $b_i=\sqrt{\lambda_i/h_i}$.

\begin{figure}[!]
\includegraphics[width=12cm]{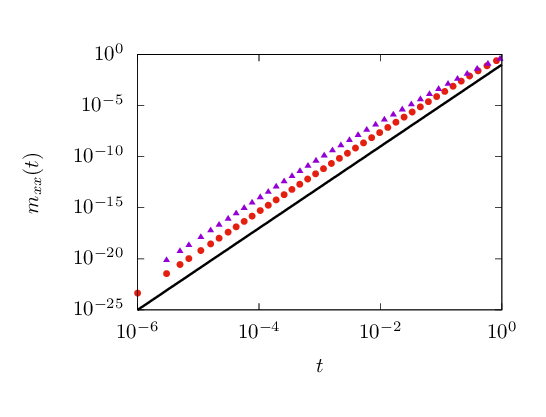}
\caption{$m_{xx}(t)$ vs $t$ for  anomalous
dissipative case $h(t)=g_\omega(t,\Gamma,N_d)$, ($k(t)=0$, $\kappa_s=0$),
with $\Gamma=10^4$ and $N_d=20$, $k(t)=0$, i.e. eq. (\ref{eqadd2}).
Symbols ($\bullet$) refer to $\omega=0.25$,
($\blacktriangle$) to $\omega=0.75$. Solid line corresponds
to $m_{xx}(t) \sim t^4$.}
\label{Fig5}
\end{figure}

\begin{figure}[!]
\includegraphics[width=12cm]{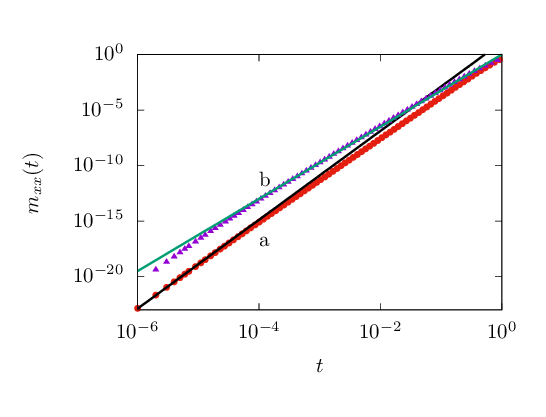}
\caption{$m_{xx}(t)$ vs $t$ for the anomalous dissipative
case $h(t)=g_\beta(t,\Gamma,N_d)$, with $k(t)=0$, $\kappa_s=0$,
eq. (\ref{eqadd2} in Appendix \ref{app2}, approximating
the local singularity of $h(t)$ near $t=0$ setting $\Gamma=10^6$.
Symbols ($\bullet$) refer to $\beta=0.25$,
($\blacktriangle$) to $\beta=0.75$. Line (a) represents
the scaling $m_{xx}(t) \sim t^4$, line (b) $m_{xx}(t)=t^{3.25}$,  fitting
the intermediate scaling of $m_{xx}(t)$ with $\beta=0.75$.}
\label{Fig6}
\end{figure}

The physical origin of a power-law dissipative kernel stems from
the highly complex rheological properties of a fluid
(as in dense polymeric fluids presenting macromolecular or particle
suspensions). In this case, while $h(0)$ is finite,
the long-term  tails of $h(t)$ may admit a power-law scaling,
$h(t) \sim t^{-\omega}$, $t \gg 1$.  For these systems,
the  short-term scaling of $m_{xx}(t)$ is unaffected
by the long-term properties of $h(t)$ and the $t^4$-scaling
is initially observed. This is depicted in figure
\ref{Fig5} considering $h(t)=g_\omega(t,\gamma,N_d)$
for a relatively ``low'' value of $\Gamma$.

A similar result occurs also if one considers diffusion in a physical
fractal
medium, in which anomalous diffusion is a consequence of
the self-similar constraints in particle motion influencing
the long-term behavior, but are immaterial for the initial dynamics of $m_{xx}(t)$ (see Appendix \ref{app2}).

Conversely, when considering a dissipative kernel $h(t)$
singular near $t=0$, as in \cite{wt,sandev_langevin_2014},
corresponding to $h(t)=  g_\beta(t,\Gamma,N_d)$ for sufficiently
high values of $\Gamma$, one recovers as an intermediate regime 
the  scaling $m_{xx}(t) \sim t^{4-\beta}$, as depicted
in figure \ref{Fig6}. Crossover between the initial
$t^4$-scaling and the intermediate scaling $m_{xx}(t) \sim t^{4-\beta}$
occurs due to the finiteness of $\Gamma$, and
the $t^{4-\beta}$-scaling sets in as the initial behavior
only in the limit for $\Gamma \rightarrow \infty$.

\section{Regularity: the rationale for non-equilibrium hydrodynamic scaling}
\label{sec5}

The entire system of articulated initial scalings
of $m_{xx}(t)$, starting from the initial
preparation $x_0=v_0=0$ (referred to as a Z-preparation, where ``Z'' stands for
zero initial conditions), that is accessible  to experimental measurement
from
the recording of a time series of the particle positions,
can  be ultimately related to the regularity of
velocity dynamics, in the meaning of H\"older regularity
of the realizations of $v(t)$.

Before addressing this issue, let us introduce
a compact interpretation of the initial non-equilibrium
scalings of the second-order moments associated with particle
observables  starting from a Z-preparation.

\subsection{Moment graph: a lumped path-integral  approach}
\label{sec5.1}

In a Z-preparation, all the moments
$m_{xx}(0)= \langle x^2(0) \rangle$,  $m_{xv}(0)= \langle x(0) \, v(0) \rangle$,
 $m_{vv}(0)= \langle v^2(0) \rangle$ - as well
the moments $m_{xq}(0)$ or $m_{vq}(0)$, in the case of
correlated patterns  (see Appendix \ref{app2}) - are vanishing, while the other degrees of
freedom associated with the memory response of the fluid are
at thermal equilibrium. Thus the resulting second-order moments attain 
finite non-zero values.
The process  that determines the initial scaling of $m_{xx}(t)$,
starting from  almost constant forcings deriving
from the equilibrium values of the remaining moments associated with
the hidden variables or from the $\delta$-correlated stochastic
forcings,  is purely a ``topological'' result
associated with the selective excitation  of the various
elements of the second-order moment hierarchy.
 
More specifically, consider the second-order moment dynamics  and define
an oriented graph in which nodes correspond to the
second-order moments of the particle observables,  with additional
nodes associated with the constant or ``almost  constant'' forcings
acting on moment dynamics. An oriented edge from node ``$\alpha$'' to
node ``$\beta$'' occurs if and only if the dynamics associated
with the moment defined by node $\beta$ contains a linear contribution
proportional to the moment associated with node $\alpha$ (or,  in the case the
node
$\alpha$ represents the constant forcing if the dynamics contains a constant
forcing).
The moment graph can be viewed as a lumped path of the observable dynamics.
Constructing the  moment graph in this way,  consider
the oriented paths from the node associated
with the constant forcing to the node associated with
the moment $m_{xx}$. The number of edges $n_e$  of the least 
path corresponds to the  exponent of the initial scaling with
time of $m_{xx}(t)$, i.e.,
\begin{equation}
m_{xx}(t) \sim t^{n_e}.
\label{eq5_1}
\end{equation}
The reason for this is rather simple: each oriented edge  from
the forcing node to the node pertaining to $m_{xx}$ corresponds to
an integration step starting from a constant forcing stepping up
to $m_{xx}$. Thus
$ \mbox{const} \mapsto t$ after one step, $\mbox{const} \mapsto  t \mapsto t^2$  
after
two steps, and so on. Therefore, the exponent $n_e$ corresponds
to the ``dimension'' of the least path connecting the
constant forcing node to  the $m_{xx}$-node.

\begin{figure}
\includegraphics[width=12cm]{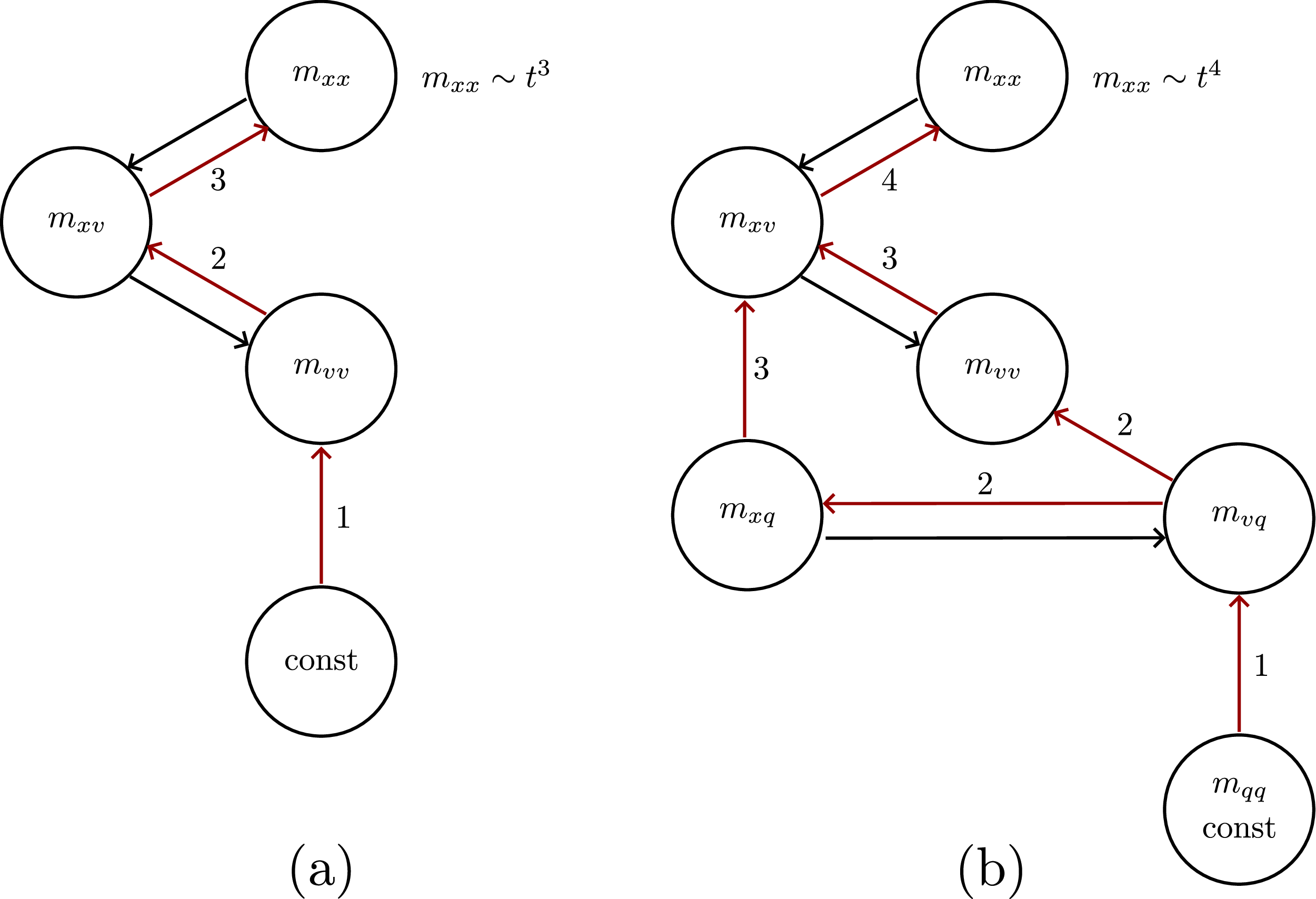}
\caption{Prototypical moment graphs associated with the Einstein-Langevin
model (panel a), 
eq. (\ref{eqadd3}), and with  its correlated fluctuational pattern (panel b), eq. (\ref{eqadd4}).}
\label{Fig7} 
\end{figure}

Figure \ref{Fig7} depicts the moment graph for that
nondimensional Einstein-Langevin model (panel a),
\begin{eqnarray}
\frac{d x}{d t} & = & v \nonumber \\
\frac{d v}{d t} & = & - v -\kappa_s \, x + \sqrt{2} \, \xi(t)
\label{eqadd3}
\end{eqnarray}
where $\xi(t)$ is the distributional derivative of a Wiener process
 and for 
correlated counterpart (panel b),
\begin{eqnarray}
\frac{d x(t)}{d t} & = & v(t) \nonumber \\
\frac{d v(t)}{d t} & = & - v(t) - \kappa_s \, x(t) + \sqrt{2} \, q(t)
\label{eqadd4} \\
\frac{d q(t)}{d t} & = & - \nu \, \left ( q(t)-\xi(t) \right ) \nonumber
\end{eqnarray}
where $\xi(t)$ acting on the momentum dynamics is substituted
by an exponentially correlated stochastic forcing $q(t)$ with correlation time $1/\nu$, $\nu>0$.
The  moment equations are reported in Appendix \ref{app2}.

Indeed, the fluid-inertial model (see eq. (\ref{eqadd1}), 
with $k(0) < \infty$) possesses
the same topological structure of the graph of the
Einstein-Langevin model, and thus at very short times $n_e=3$.
Similarly, all the correlated fluctuational patterns associated
with Brownian dynamics possess the same topological
structure of the graph in panel (b), and thus $n_e=4$.
In the same class falls the graph of the viscoelastic
model in the absence of fluid inertial effects, for which $n_e=4$.
Of course, this analysis applies to the cases where the memory kernels 
do not possess a singularity at $t=0$, and therefore this lumped approach cannot recover the intermediate scaling
$m_{xx}(t) \sim t^{5/2}$ typical of the intermediate $1/\sqrt{t}$ scaling
of the Basset fluid-inertial kernel.

But even in this case, the graph approach provides a general
and valuable result: for an initial Z-preparation ($x_0=v_0=0$), if
$m_{vv}(t) \sim t^\zeta$, then $m_{xx}(t) \sim t^{2+\zeta}$.
The usefulness of this elementary observation that follows
straightforwardly from the structure of the moment graph
is addressed in the next paragraph.

\subsection{Regularity of the Brownian velocity $v(t)$}
\label{sec5.2}

The initial scaling of $m_{xx}(t)$ starting from a Z-preparation is
the manifestation of the regularity properties of the particle
velocity or, more precisely, of its realizations, viewed as
 functions of the time variable $t$.
A realization  $v(t)$  of the particle velocity in a generic 
time interval $t \in [0,t_{\rm max}]$, $t_{\rm max}>0$, is a H\"older continuous
function of exponent $H_v$ if, for any $t_1, \, t_2>t_1$, $t_1, \, t_2 \in [0,t_{\rm max}]$,
there exist positive constants $c_1$, $c_2 > c_1$, such that
\begin{equation}
c_1 \, | t_2 - t_1 |^{H_v} \leq | v(t_2)-v(t_1) | \leq c_2 \,  | t_2 - t_1 |^{H_v}  \, .
\label{eq5_2}
\end{equation}
If $H_v=1$, then $v(t)$ is Lipschitz continuous, while
a H\"older exponent $H_v<1$  for any interval $[t_a,t_b]$, with $t_b>t_a$,
 implies that the graph of $v(t)$ is
almost everywhere a non-differentiable curve possessing fractal
dimension $d_v$ given by
\begin{equation}
d_v = 2 - H_v  \, .
\label{eq5_3}
\end{equation}
The particle  velocity is a stochastic process and thus eq. (\ref{eq5_2})
applies to a generic realization of it almost everywhere with
probability $1$.
The same bounds apply to the  expected value 
$\langle | v(t_2)-v(t_1) | \rangle$
with respect to the probability measure of the  velocity fluctuations.

In  the framework of regularity analysis, the initial
scaling of $m_{vv}(t)$ starting from a Z-preparation
provides an alternative way to estimate
the H\"older exponent $H_v$, as
\begin{equation}
\langle | v(t_2)-v(t_1) | \rangle_{\rm eq} \sim
\left \langle ( v(t_2)-v(t_1)  )^2\right \rangle_{\rm eq}^ {1/2} = \sqrt{m_{vv}(t)}
\label{eq5_4}
\end{equation}
where $\langle \cdot \rangle_{\rm eq}$ indicates that
the expected values refer to the analysis of an equilibrium
realization of the Brownian dynamics.

It follows from eq. (\ref{eq5_4}) that the exponent $\zeta$ controlling
the initial scaling of  $m_{vv}(t) \sim t^\zeta$ (starting from a  Z-preparation)
is related to the H\"older exponent $H_v$ via the relation
\begin{equation}
H_v= \frac{\zeta}{2}
\label{eq5_5}
\end{equation}
Of course, this analysis applies at short time scales
and this poses the practical experimental problem of measuring
$v(t)$ at very short time-scales. The non-equilibrium
analysis starting from a Z-preparation provides an efficient way
to circumvent this problem  by considering the scaling of $m_{xx}(t)$, a
 quantity more accurate  and robust to  determine experimentally.
From the observation addressed at the end of paragraph \ref{sec5.1}
(i.e. if $m_{vv}(t) \sim t^\zeta$, then $m_{xx}(t) \sim t^{2+\zeta} = t^\varphi$),
the exponent $\varphi$ controlling
the initial scaling of $m_{xx}(t)$
is related to the H\"older exponent $H_v$  of the particle velocity 
via the equation 
\begin{equation}
H_v= \frac{\varphi -2}{2}  \, .
\label{eq5_7}
\end{equation}
Eq. (\ref{eq5_7}) represents a powerful  relation to assess experimentally
the scaling controlling Brownian motion fluctuations in a fluid medium.
Indeed, depending on the regularity of the particle velocity, different
universality classes of Brownian motion fluctuations can be
defined, the occurrence of which depends on the hydrodynamic regimes,
i.e. on the specific fluid-particle interactions at thermal equilibrium.
Table \ref{Tab1} reviews the value of the scaling exponents characterizing
non-equilibrium Brownian  dynamics in a harmonic trap for different 
considered models of increasing hydrodynamic complexity.

\begin{table}[ht]
\centering
\begin{tabular}{cccc}
\hline  \hline
Model & $ \quad \varphi \quad $ & $ \quad H_v \quad $ & $\quad d_v \quad $ \\
\hline
EL model eq. (\ref{eqadd3}) & $3$ & $1/2$ & $3/2$ \\
FI effects eq. (\ref{eqadd1}), $k(0) \rightarrow \infty$  & $5/2$ & $1/4$ & $7/4$ \\ 
FI effects eq. (\ref{eqadd1}), $k(0) < \infty$  & $3$ & $1/2$ & $3/2$ \\
Viscoelastic case eq.  (\ref{eqadd2}) no FI & $4$ & $1$ & $1$ \\
Viscoelastic case eq.   with FI $k(0) < \infty$  & $3$ & $1/2$ & $3/2$ \\
Correlated patterns eq.   (\ref{eqadd4}) & $4$  & $1$ & $1$ \\
Anomalous dissipative model eq. (\ref{eqadd2}), $h(0) < \infty$ & $4$ & $1$ & $1$ \\
\hline \hline
\end{tabular}
\caption{Review of the scaling exponents characterizing initial Brownian
motion, starting from a Z-preparation for the different
hydromechanic models. EL stands for Einstein-Langevin, FI for fluid inertia.
The case $k(0) \rightarrow \infty$ corresponds to the singular 
Basset kernel.}
\label{Tab1}
\end{table}
\begin{figure}
\includegraphics[width=12cm]{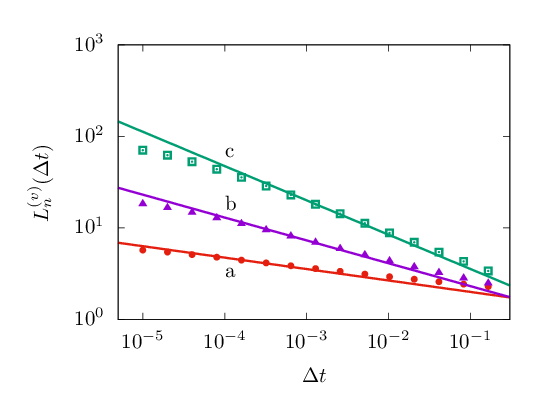}
\caption{$L_n^{(v)}(\Delta t)$ vs $\Delta t$  for the singular
dissipative dynamics characterized by the memory kernel
$h(t)=g_\beta(t,\Gamma,N_d)$ with $\Gamma=10^6$, $N_d=20$,
while $k(t)=0$ and $\kappa_s=0$, eq. (\ref{eqadd2}). 
 Symbols ($\bullet$)  refer to $\beta=0.25$,
($\blacktriangle$) to $\beta=0.5$, ($\square$) to $\beta=0.75$.
The solid lines (a) to (c) correspond to the scaling
laws $L_n^{(v)}(\Delta t) \sim \Delta t^{-\beta/2}$
for the three different values of $\beta$.}
\label{Fig8} 
\end{figure}

To make an example, consider the purely dissipative memory dynamics
controlling by the kernel $h(t)$ possessing a singularity at $t=0$,
$h(t) \sim t^{-\beta}$, with $0 <\beta <2$, while $k(t)=0$.
In this case from eqs. (\ref{eq5_3}) and (\ref{eq5_7}), it follows
that the fractal dimension of the velocity process equals $d_v=1+\beta/2$.
The fractal dimension $d_v$ can be obtained
from the length-resolution analysis \cite{frac1,frac2}
of a generic realization of the velocity process in equilibrium
conditions, by considering the normalized length  $L_n^{(v)}(\Delta t)$ of the graph of
a realization of $v(t)$ over the interval $[0,t_{\rm max}]$ rescaled by
$t_{\rm max}$,
\begin{equation}
L_n^{(v)}(\Delta t) \sim \Delta t^{-(1-d_v)}  \, .
\label{eqlength}
\end{equation}
In the present case, $L_n^{(v)}(\Delta t)  \sim \Delta t^{-\beta/2}$.
Figure \ref{Fig8} depicts $L_n^{(v)}(\Delta t)  $ vs $\Delta t$ in this
case, where $h(t)=g_\beta(t,\Gamma,N_d)$ with $\Gamma=10^6$, $N_d=20$,
for different values of $\beta$, where $t_{\rm max}=1.67 \times 10^2$.
Since $h(0) <\infty$, due to the finite value of $\Gamma$,
the fractal scaling eq. (\ref{eqlength}) applies to intermediate
time scales, as in the very early regime $\beta=0$, $\varphi=4$ and thus
$d_v=1$.

\begin{table}[ht]
\centering
\begin{tabular}{c|c|c|c|c|c|c}
\hline  \hline
$\quad$ Scaling $\quad$ &  $\,$ Exponent $\,$ & $ \quad \beta \quad $ & $ \quad \zeta \quad $ & $\quad \varphi \quad $& $ \quad H_v \quad $ &  $\quad d_v \quad $ \\
\hline
 $\quad H(t) \sim t^{-\beta} \quad $ & $\beta$ & $-$ & $2-\zeta$  & $4 -\varphi$  &
$\; 2- 2 \, H_v\; $ & $\;2 \, d_v -2\; $ \\
$\quad m_{vv}(t) \sim t^\zeta \quad $ & $\zeta$ & $2-\beta$ & $-$ & $\varphi-2$ & $2 \, H_v$
& $4 -2 \, d_v$ \\
$\quad m_{xx}(t) \sim t^\varphi \quad$ &  $\varphi$  & $4-\beta$ & $2+\zeta$ & $-$ &
$2 + 2 \, H_v$ & $6-2 \, d_v$ \\
$\langle | v(t+\Delta t) - v(t) | \rangle_{\rm eq} \sim \Delta t^{H_v} \;$ & 
$H_v$ & $\; (2-\beta)/2 \; $ & $\zeta/2$ & $\; (\varphi-2)/2 \;$ & $-$ & $2-d_v$ \\
$L_n^{(v)}(\Delta t) \sim \Delta t^{-(1-d_v)}$ & $d_v$  &
$\; (\beta+2)/2 \;$ & $\; (4-\zeta)/2\; $ & $\; (6-\zeta)/2\;$ & $2-H_v$ & $-$ \\
\hline \hline
\end{tabular}
\caption{Synopsis of the main scaling laws involved in the short-time/regularity
analysis of Brownian motion experiments starting from a Z-preparation, and
of the relations amongst the corresponding scaling exponents.}
\label{Tab2}
\end{table}
Table \ref{Tab2} reports a synopsis of the scaling theory presented in
this paragraph with regards to the main quantities involved in the
short-time dynamics and in the regularity analysis of Brownian fluctuations.
It  should be observed that the relations amongst the exponents
reported in Table \ \ref{Tab2} hold for $\beta<2$.
At $\beta=2$, the value of $d_v$ equals $2$, i.e. the curve
of the velocity realizations ``fill the plane'', and this
represents a limit condition.
 If $\beta>2$, the relations
amongst the exponents
in Table II still apply by substituting the exponent 
$\beta$ with $\mbox{min}(\beta,2)$.

\subsection{Comparison with experimental results}
\label{sec5.3}

There is a broad evidence of experimental results supporting 
the  non-equilibrium Brownian motion theory developed in the previous
paragraphs.
The work by Duplat et al. \cite{duplat} has shown that $\varphi=3$ for
Brownian motion in gases, consistent with the properties
that in gases fluid-inertial effects are negligible
(the difference of the particle to fluid density is of three
order of magnitude) and thus the Einstein-Langevin model (for the
time scales considered in the experiments) is reasonably valid.

In liquids,  a  recent work \cite{btr}
has  shown experimentally
the $5/2$-law for $\varphi$ in liquids deriving
from the inertial effects, due to
the singular nature of the  Basset kernel.
The time scales considered in \cite{btr}  do not allow the possible
observation of an earlier $t^4$-scaling, associated with
the  regular nature of the velocity fluctuations.
In this hydrodynamic regime, and in this range of time scales, the
fractal nature of  experimental Brownian velocity realizations in
liquids is addressed in \cite{noifrac1}
 where,  via length-resolution analysis
the  fractal dimension $d_v=7/4$  for Brownian velocity fluctuations
is recovered. This is a further
experimental validation of the importance of fluid-inertial
effects, modifying the universality class of Brownian motion.

\section{Concluding remarks}
\label{sec6}
We have provided the theoretical  analysis of non-equilibrium
trapped Brownian  dynamics and of its scaling properties.
The founding principle  is that
an  equilibrium trajectory contains
information about non-equilibrium conditions  of the
particle, that can be recovered by applying a
selective subsampling of the  equilibrium trajectory itself.
This is essentially a consequence of the Chapman-Kolmogorov
equation and of its extension to non-Markovian cases admitting
a Markovian embedding.
This approach opens up the possibility of deepening our
understanding of Brownian motion beyond the pure equilibrium
description. 

Particularly interesting is the Z-preparation,  corresponding
to the subsampling of trajectory portions where initially
both the velocity and the particle position are vanishing (in the
meaning that the origin for the particle position   corresponds
to its equilibrium point in the harmonic trap).
The analysis has been carried out following two
distinct but complementary approaches: the
free initial process approach based on eq. (\ref{free_inertial}),
and the moment analysis starting from the Markovian
embedding of the memory effects, based on the fluctuational
patterns introduced in \cite{gpp1} and reviewed in Appendix
\ref{app2}. While eq. (\ref{free_inertial}) has been previously
proposed in \cite{wm}, its application to
include inertial effects, and its practical feasibility
in analyzing experimental time series  in highly resolved
Brownian measurements represents a major result
in Brownian motion theory, as it makes the 
analysis of short-time non-equilibrium conditions
amenable to experimental validation.

The  main results of the article can be summarized as follows:
\begin{itemize}
\item The initial/intermediate scaling, particularly the
scaling exponent $\varphi$ for $m_{xx}(t)$ in Z-prepared conditions,
yields  qualitative information on the hydrodynamic fluid
regimes.
\item The exponent $\varphi$ in Z-prepared conditions is
related to the H\"older regularity exponent of the velocity
fluctuations of the Brownian particle (see eq. (\ref{eq5_7})).
\item   A complete scaling theory  has been derived connecting
the short-time dynamic  and the regularity
exponents of Brownian velocity fluctuations 
(see Table \ref{Tab2})
in  different hydrodynamic and rheological conditions.
\end{itemize}

This means that a  Brownian  particle can be used as a probe
to investigate the hydrodynamic properties of the solvent
fluid in which it is immersed, 
as different
memory effects induced by the properties of the
fluid (viscoelasticity and fluid inertia)  induce
completely different statistical properties
in non-equilibrium Brownian motion dynamics, both in
the initial and the intermediate scalings of $m_{xx}(t)$.

The connection between early-intermediate scaling properties 
and the regularity of particle velocity  deserves attention.
It has been shown, consistent with the analysis
reported in \cite{btr}, 
that a strict singular Basset kernel provides a $t^{5/2}$-scaling
for the Z-prepared $m_{xx}(t)$.
As discussed elsewhere, the strict $1/\sqrt{t}$ singularity
of the fluid inertial kernel is a mathematical approximation
stemming from the assumption of infinite velocity of propagation
of the shear stresses. The inclusion of viscoelastic effects
removes this singularity. Correspondingly,  if one adheres
to the classical Kubo theory, one should observe the
early  $t^3$-scaling of $m_{xx}(t)$ followed by the
inertial $t^{5/2}$-scaling.  If one considers water or acetone as 
the
solvent fluid, for which the characteristic relaxation times
are order of $1$ ps, the crossover time
between these
two $m_{xx}(t)$ scaling regimes is more than four order of magnitude lower than
the momentum relaxation time of a micrometric Brownian particle. 
This is far beyond the actual reach of experimental resolution.

Conversely, within the Kubo fluctuation-dissipation
formulation, if a finite
correlation time is included in the hydrodynamic forcings acting on the
memory degrees of freedom of the particle hydromechanics,
an initial $t^4$-scaling of $m_{xx}(t)$ should be observed,
owing to the regularity of particle velocity realizations ($H_v=1$).

As a final remark, the results presented in this
article have shown that the regularity
properties  of Brownian motion, associated
with the scaling of $\langle |v(t+\Delta t)-v(t)| \rangle_{\rm eq}$
or of the length $L_n^{(v)}(\Delta t)$ of a particle
velocity realization vs the temporal resolution $\Delta t$, are
not only physical quantities amenable to experimental
measurements in highly resolved experiments (see \cite{noifrac1}),
but  also possess physical relevance comparable to
the lower-order moments $m_{xx}(t)$ and $m_{vv}(t)$
of the particle position and velocity.
The  relations amongst the exponents associated with
these two classes
of quantities (Table \ref{Tab2}) connect equilibrium
properties, namely $\langle |v(t+\Delta t)-v(t)| \rangle_{\rm eq}$
and  $L_n^{(v)}(\Delta t)$, to non-equilibrium dynamics
expressed by $m_{xx}(t)$ or $m_{vv}(t)$ starting from
a Z-preparation. This is a further consequence of
the superposition principle of non-equilibrium states
within an equilibrium trajectory, stated in section \ref{sec2}.

\appendix
\section{On the structure of inertial kernels}
\label{app1}

The fluid inertial contribution defined
by the kernel $k(t)$ in eq. (\ref{eq2_1}) 
can be correctly transformed into a global
kernel $H(t)$ solely if $k(0) <\infty$.
In this case, from eq. (\ref{eq2_1}) we have
\begin{equation}
- \int_0^t k(t-\tau) \, \left ( \frac{d v(\tau)}{d \tau} + v(0) \, \delta(\tau)
\right ) \, d \tau = -k(0) \, v(t)
+\int_0^t \frac{d k(t-\tau)}{d \tau} \, v(\tau) \, d \tau \, ,
\label{eq_az1}
\end{equation}
thus obtaining
\begin{equation}
H(t)=  h(t)+k(0) \, \delta(t) + \frac{d k(t)}{d t} \, .
\label{eq_az2}
\end{equation}
If $k(0)$ is unbounded, the kernel $H(t)$ cannot
be defined, and similarly the stochastic force $R(t)$.
The kernel $H(t)$  defined by eq. (\ref{eq_k2}) implies
$h(t)=\eta \, \delta(t)$ and $k(0)=0$, and the latter condition
displays inconsistencies impacting on the long-term dynamics.
For instance, it does not satisfy dissipative properties.
By definition, the global friction factor $\eta_\infty$ in
the presence of memory dynamics entering the global
Stokes-Einstein fluctuation-dissipation relation involving particle
diffusivity $D$, 
\begin{equation}
D \, \eta_\infty = k_B \, T \, ,
\label{eq_az3}
\end{equation}
 is
the integral of $H(t)$,
\begin{equation}
\eta_\infty = \int_0^\infty H(t) \, d t \,  .
\label{eq_az4}
\end{equation}
In the case of eq. (\ref{eq_k2}) one obtains,
\begin{equation}
\int_0^\infty \left ( \eta \, \delta(t) - \frac{b}{2 \, t^{3/2}}
\right ) \, d t = -\infty
\label{eq_z5}
\end{equation}
which is manifestly unphysical, since it is diverging
(and this is not the case for a Brownian
particle in a Newtonian fluid) and also negative (and this
goes against any  thermodynamic principle of dissipation).
Therefore, the kernel $H(t)$ defined by eq. (\ref{eq_k2})
fails to describe the long-term properties of Brownian
dynamics, and cannot be used to model Brownian
dynamics in equilibrium conditions.

Despite these conceptual shortcomings, eq. (\ref{eq_k2})
still represents an accurate approximation of the short-time
behavior starting from $t>t_\lambda$, where $t_\lambda$
is of the order of magnitude of the characteristic
relaxation time (for water order of $1$ ps).
This is because eq. (\ref{eq_k2}) overlooks
the term $k(0) \, \delta(t)$ that provides a
contribution of higher-order ($t^3$)  to $m_{xx}(t)$
with respect to the inertial term that gives rise to
the $t^{5/2}$-scaling.
Moreover, for $t>t_\lambda$, the  continuous
part at the r.h.s. of eq. (\ref{eq_k2}) represents
a good approximation of the term $dk(t)/dt$ entering eq.
(\ref{eq_az2}).
With this caveat, eq. (\ref{eq_k2}) can be properly used
in the free inertial process formalism  developed in
paragraph \ref{sec4_1} to derive qualitatively and quantitatively
the short-time scaling of $m_{xx}(t)$.

\section{Modal analysis of hydrodynamic regimes}
\label{app2}

This appendix reviews and makes explicit the formalism
underlying modal expansion of Brownian motion dynamics
in different hydrodynamic regimes consistent with the
Kubo fluctuation-dissipation theory   and some
auxiliary results of this analysis.
 
To begin with, consider the simplest case of a spherical free particle
of mass $m$
in a solvent fluid at constant temperature $T$
in the presence of instantaneous fluid-particle
interactions, i.e. in the Stokesian regime, confined by
a harmonic optical trap of spring constant $k_s$,
\begin{eqnarray}
\frac{d x(t)}{d t} & = & v(t) \nonumber \\
m \, \frac{d v(t)}{d t} & = &- \eta \, v(t)- k_s \, x(t) + \sqrt{2 \, k_B \, T \, \eta}
\, \xi(t) 
\label{eq4_1}
\end{eqnarray}
where $\eta$ is the particle friction factor and
$\xi(t)=d w(t)/dt$ is the distributional
derivative of a one-dimensional Wiener process $w(t)$.
This represents the classical Einstein-Langevin model, that in
the presence of a harmonic potential is customarily
referred to as an Orstein-Uhlenbeck process.
In dimensionless terms, rescaling time by the momentum relaxation
time $T_c=\tau_p=m/\eta$, and velocity by the equilibrium thermal
velocity $V_c$,
\begin{equation}
V_c=\sqrt{\frac{k_B \, T}{m}}
\label{eq4_2}
\end{equation}
so that the characteristic length $L_c$ becomes
\begin{equation}
L_c= V_c \, T_c = \sqrt{\frac{k_B \, T  \, m}{\eta^2}}
\label{eq4_3}
\end{equation}
the equations of motion  in nondimensional form,
(keeping the same symbols for the nondimensional quantities),
become
\begin{eqnarray}
\frac{d x}{d t} & = & v \nonumber \\
\frac{d v}{d t} & = & - v -\kappa_s \, x + \sqrt{2} \, \xi(t)
\label{eq4_8}
\end{eqnarray}
where  the dimensionless spring constant $\kappa_s$ is related
to the spring constant $k_s$ [N/m] of the trap by the relation
$\kappa_s=k_s \, m/\eta^2$.
The associated Fokker-Planck equation for the probability
density $p(x,v,t)$ reads
\begin{equation}
\frac{\partial p}{\partial t}= - v \, \frac{\partial p}{\partial x}
+ \frac{\partial (v \, p)}{\partial v}+ \kappa_s \, x \, \frac{\partial p}{\partial v}+ \frac{\partial^2 p}{\partial v^2}  \, .
\label{eq4_9}
\end{equation}
For a trapped particle, the second-order moment equations read
\begin{eqnarray}
\frac{d m_{xx}}{d t} & = & 2 \, m_{x v} \nonumber \\
\frac{d m_{xv}}{d t} & =  & m_{vv}-m_{x v} - \kappa_s \, m_{xx}
\label{eq4_15} \\
\frac{d m_{vv}}{d t} & = & - 2 \, m_{vv} -2 \, \kappa_s \, m_{x v}
+2 
\nonumber 
\end{eqnarray}
Consider the initial Z-preparation
 $x_0=0$, $v_0=0$,
leading to the analysis of $m_{xx}(t)=m_{xx}(t \, | \, x_0=0, v_0=0)$.
In this particular subsampling, the  initial condition for the 
moments are
$m_{xx}(0)=m_{xv}(0)=m_{vv}(0)=0$,
and we have  at short time scales
\begin{equation}
m_{xx}(t) = \frac{2}{3} \, t^3+ O(t^4) \, .
\label{eq4_16}
\end{equation}

The  same analysis can be extended to interactions characterized
by memory effects.

\subsubsection{Viscoelastic fluid}
\label{sec4.2a}
To begin with, consider the case of a Maxwell fluid characterized by a single
relaxation rate $\lambda$. This case has been considered in \cite{wm}.

The nondimensional modal expansion of particle
hydromechanics, consistent with fluctuation-dissipation relations, and
  including
also the presence of a harmonic trap is,
\begin{eqnarray}
\frac{d x}{d t} & = & v \nonumber \\
\frac{d v}{d t}  & =  & - \lambda \, \theta - \kappa_s \, x
\label{eq4_21} \\
\frac{d \theta}{d t} & = & - \lambda \, \theta + v + \sqrt{2} \, \xi(t)
\nonumber
\end{eqnarray}
The setting is such that the long-term nondimensional friction
coefficient equals $1$.
In this case, the Fokker-Planck equation for the probability density $p(x,v,\theta, t)$ reads
\begin{eqnarray}
\frac{\partial p}{\partial t}  & = & - v \, \frac{\partial p}{\partial x}+
\lambda \, \theta \, \frac{\partial p}{\partial v} +\kappa_s \, x \, \frac{\partial p}{\partial v}+ \lambda \, \frac{\partial (\theta \, p )}{\partial \theta}
 -  v \, \frac{\partial p}{\partial \theta} + \frac{\partial^2 p}{\partial \theta^2}
\label{eq4_22}
\end{eqnarray}
from which the dynamics of the second-order moments can be derived
\begin{eqnarray}
\frac{d m_{xx}}{d t} & = & 2 \, m_{x v} \nonumber \\
\frac{d m_{xv}}{d t} & = &  m_{vv} - \lambda \, m_{x \theta} -\kappa_s \, m_{xx} \nonumber \\
\frac{d m_{x \theta}}{d t} & = & m_{v \theta} - \lambda \, m_{x \theta}+ m_{xv}
\nonumber \\
\frac{d m_{vv}}{d t} & = &  - 2 \, \lambda \, m_{v \theta} -2 \, \kappa_s \, m_{xv}  \label{eq4_23} \\
\frac{d m_{v \theta}}{d t} & = & - \lambda \, m_{\theta \theta} - \kappa_s \,
m_{x \theta} - \lambda \, m_{v \theta} + m_{vv}
\nonumber \\
\frac{d m_{\theta \theta }}{d t} & = &  - 2 \, \lambda \, m_{\theta \theta}
+ 2 \, m_{v \theta} +2
\nonumber
\end{eqnarray}
The equilibrium conditions (both for the medium and the
particle) correspond  to
\begin{eqnarray}
m_{xx}^* & =  &
\frac{1}{\kappa_s} \,, \quad m_{vv}^*= 1 \, , \quad m_{\theta \theta}^*=\frac{1}{\lambda} \,, \quad
m_{xv}^*  =  m_{x \theta}^*=m_{v \theta}^* =0
\label{eq4_24}
\end{eqnarray}
 Consider a Z-prepared initial state and equilibrium
conditions for the fluid, i.e. the memory degrees of freedom
 of the fluid are at their equilibrium values.
 In this case, since $x_0=v_0=0$, the initial
conditions for the moments are
\begin{equation}
m_{xx}(0)=m_{xv}(0)=m_{x \theta}(0)=m_{v v}(0)=m_{v \theta}(0) \, ,
\quad m_{\theta \theta}(0)=m_{\theta \theta}^*=1/\lambda \, .
\label{eq4_25}
\end{equation}
A good approximation  in the early stages of the dynamics is to
assume
\begin{equation}
m_{\theta \theta}(t) \simeq m_{\theta \theta}^* = \mbox{const}
\label{eq4_26}
\end{equation}
from which the following scalings follow
\begin{equation}
\frac{d m_{v \theta}}{d t} \simeq - \lambda m_{\theta \theta}^*
\quad \Rightarrow \quad m_{v \theta}(t) \simeq -\lambda \, m_{\theta \theta}^* \,t
\label{eq4_27}
\end{equation}
\begin{equation}
\frac{d m_{x \theta}}{d t} \simeq  m_{v \theta}
\quad \Rightarrow \quad m_{x \theta}(t) \simeq -\frac{\lambda \, m_{\theta \theta}^*}{2} \,
t^2
\label{eq4_28}
\end{equation}
\begin{equation}
\frac{d m_{v v}}{d t} \simeq - 2 \, \lambda m_{v \theta}  \simeq
2 \, \lambda^2 m_{\theta \theta}^* \, t
\quad \Rightarrow \quad m_{v v}(t) \simeq \lambda^2 \, m_{\theta \theta}^* \,
t^2
\label{eq4_29}
\end{equation}
\begin{equation}
\frac{d m_{x v}}{d t} \simeq m_{vv}- \lambda m_{x \theta} =  \frac{3 \,
\lambda^2 m_{\theta \theta}^*}{2} \, t^2
\quad \Rightarrow \quad m_{x v}(t) \simeq \frac{\lambda^2 \, m_{\theta \theta}^*}{2} \,
t^3
\label{eq4_30}
\end{equation}
\begin{equation}
\frac{d m_{x x}}{d t} = 2 \,  m_{xv}  \simeq
\lambda^2 m_{\theta \theta}^* \, t^3
\quad \Rightarrow \quad m_{x x}(t) \simeq \frac{\lambda^2 \, m_{\theta \theta}^*}{4} \,
t^4
\label{eq4_31}
\end{equation}
Therefore, in a viscoelastic medium (we are considering a Maxwell
fluid, but this result can be  generalized to any linear viscoelastic
response, see eq. (\ref{eq_d1}))
 the initial dynamics of $m_{xx}(t)$
are characterized by a power-law scaling with an exponent
equal to 4.

\subsubsection{Fluid-inertial effects}
\label{sec4.2b}
In this paragraph we consider the effect of fluid inertia, i.e.
of the Basset force.
The fluctuational pattern for a Brownian
particle in a harmonic trap  in the
presence of an instanteneous Stokesian friction and  fluid inertial
effects is expressed by
\begin{eqnarray}
\frac{d x(t)}{d t} & = & v(t) \nonumber \\
\frac{ d v(t)}{d t} & = & - v(t) - \kappa_s \, x(t)- G \, v(t) +
\sum_{\alpha=1}^{N_i} \gamma_\alpha \, \mu_\alpha \, z_\alpha(t)
+ \sqrt{2} \,  \sum_{\alpha=1}^{N_i} d_\alpha \, \xi_\alpha(t) + \sqrt{2} \, \xi(t) \label{eqi_1} \\
\frac{d z_\alpha(t)}{d t} & = & - \mu_\alpha \, z_\alpha(t) + v(t) + \sqrt{2}
\, c_\alpha\, \xi_\alpha(t) \nonumber
\end{eqnarray}
where  $\xi(t)$ and $\xi_\alpha(t)$, $\alpha=1,\dots,N_i$
are distributional derivatives of independent Wiener processes,
$G=\sum_{\alpha=1}^{N_i} \gamma_\alpha$, and
the coefficients $d_\alpha$ and $c_\alpha$ modulating the ``inertial''
noise take the values \cite{gpp1}
\begin{equation}
d_\alpha= \sqrt{\gamma_\alpha} \, , \quad c_\alpha= - \frac{1}{\sqrt{\gamma_\alpha}} \, , \quad \alpha=1,\dots,N_i  \, .
\label{eqi_2}
\end{equation}
Observe also the presence of the other fluctuational
contribution $\sqrt{2} \, \xi(t)$ in the velocity equation,
 independent of the
forcings $\xi_\alpha(t)$, occurring due to the presence of
the instantaneous Stokes friction $-v(t)$.
No hydrodynamic added-mass effects are included. Accordingly, $k(0)$ is finite,
consistent with a dissipative kernel characterized by a non-vanishing
shear-stress relaxation time.

In this model the fluid-inertial kernel can be taken
proportional to the function $g(t,\Gamma,N_i)$, i.e.
$k(t)= b \, g(t,\Gamma,N_i)$, i.e eq. (\ref{eqi_3}) with $\xi=1/2$,
and therefore it can be
expressed as
$\sum_{\alpha=1}^{N_i} \gamma_\alpha \, e^{-\mu_\alpha t}$.

Due to the coupling between the dynamics of $v(t)$ and $z_\alpha(t)$,
as regards the fluctuational forcings $\xi_\alpha(t)$,
the Fokker-Planck equation for the probability density reads
\begin{eqnarray}
\frac{\partial p}{\partial t} & = & - v \, \frac{\partial p}{\partial x}
+ \frac{\partial (v \, p)}{\partial v} + \kappa_s \, x  \, \frac{\partial p}
{\partial v} + G \, \frac{\partial (v \, p)}{\partial v}
- \sum_{\alpha=1} \gamma_\alpha \, \mu_\alpha \, z_\alpha \frac{\partial p}{\partial v} + \sum_{\alpha=1}^{N_i} \frac{\partial (\mu_\alpha \, z_\alpha \, p)}{\partial z_\alpha}\nonumber \\
&- &  \sum_{\alpha=1}^{N_i} \frac{\partial (v \, p)}{\partial z_\alpha}
+ \sum_{\alpha=1}^{N_i} d_\alpha^2 \, \frac{\partial^2 p}{\partial v^2}
+ 2 \sum_{\alpha=1}^{N_i} c_\alpha \, d_\alpha \, \frac{\partial^2 p}{\partial v \partial z_\alpha}+ \sum_{\alpha=1}^{N_i} c_\alpha^2 \, \frac{\partial^2 p}{\partial z_\alpha^2} + \frac{\partial^2 p}{\partial v^2}
\label{eqi_6}
\end{eqnarray}
from which the moment equations follow
\begin{eqnarray}
\frac{d m_{xx}}{d t} & = & 2  \, m_{x v} \nonumber \\
\frac{d m_{x v}}{d t} & = & m_{vv} - m_{x v}- \kappa_s \,m_{xx}
- G \, m_{x v}+ \sum_{\beta=1}^{N_i} \gamma_\beta \, \mu_\beta \,
m_{x z_\beta} \nonumber \\
\frac{d m_{x z_\alpha}}{d t} & = &  m_{v z_\alpha}- \mu_\alpha \, m_{x z_\alpha}
+ m_{xv} \label{eqi_7}  \\
\frac{d m_{vv}}{d t} & = & - 2 \, m_{vv}- 2 \, \kappa_s \, m_{x v} -
2 \, G \, m_{vv}+ 2 \, \sum_{\beta=1}^{N_i}
\gamma_\beta \, \mu_\beta \,m_{v z_\beta}
+ 2 \, \sum_{\beta=1}^{N_i} d_\beta^2 +2  \nonumber \\
\frac{d m_{v z_\alpha}}{d t} & = &  - m_{v z_\alpha} -
\kappa_s \, m_{x z_\alpha} - G \, m_{v z_\alpha}+
\sum_{\beta=1}^{N_i} \gamma_\beta \, \mu_\beta \, m_{z_\alpha z_\beta}
- \mu_\alpha \, m_{v z_\alpha}+ m_{vv}+ 2 \, c_\alpha \, d_\alpha
\nonumber \\
\frac{d m_{z_\alpha z_\beta }}{d t} & = &  - (\mu_\alpha+\mu_\beta) \,
+m_{v z_\alpha} + m_{v z_\beta} +2 \, c_\alpha^2 \, \delta_{\alpha \beta}
\nonumber
\end{eqnarray}
The total number of moment equations is $3+ 2 N_i+ N_i^2$.

\begin{figure}
\includegraphics[width=12cm]{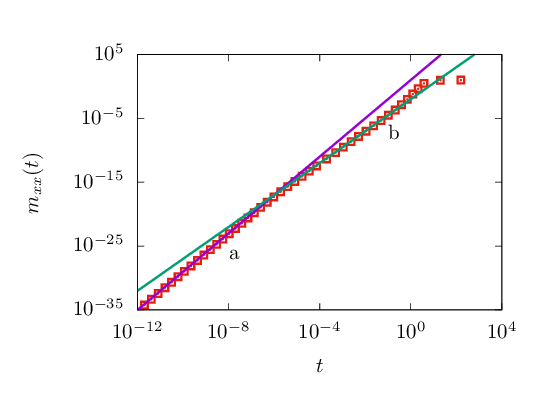}
\caption{$m_{xx}(t)$ vs $t$ obtained from eq. (\ref{eqi_9}), i.e.
for a Maxwell fluid in the presence of fluid inertial effects, at
$\kappa_s=0.1$, $b=0.1$ for different values of $\lambda=0.1, \, 1,\, 10$
(symbols).  The data for different $\lambda$'s overlap and are practically
indistinguishable in a log-log plot.
Line (a) represents the scaling $m_{xx}(t) \sim t^3$,
line (b) $m_{xx}(t) \sim t^{5/2}$.}
\label{Fig_a1}
\end{figure}

Before discussing the scaling behavior of the second-order moments in
a Newtonian fluid, let us
consider the same problem in a presence of a Maxwell fluid
characterized by a single relaxation rate $\lambda>0$ defined
by the  equations of motion
\begin{eqnarray}
\frac{d x(t)}{d t} & = & v \nonumber \\
\frac{ d v(t)}{d t} & = & - \lambda \, \theta(t) - \kappa_s \, x(t)- G \, v(t) +
\sum_{\alpha=1}^{N_i} \gamma_\alpha \, \mu_\alpha \, z_\alpha(t)
+ \sqrt{2} \,  \sum_{\alpha=1}^{N_i} d_\alpha \, \xi_\alpha(t)  \label{eqi_8} \\
\frac{d \theta(t)}{d t} & = & - \lambda \, \theta(t) + v(t) +
\sqrt{2} \, \xi(t) \nonumber \\
\frac{d z_\alpha(t)}{d t} & = & - \mu_\alpha \, z_\alpha(t) + v(t) + \sqrt{2}
\, c_\alpha\, \xi(t)  \nonumber
\end{eqnarray}
from which the moment equations follow
\begin{eqnarray}
\frac{d m_{xx}}{d t} & = & 2  \, m_{x v} \nonumber \\
\frac{d m_{x v}}{d t} & = & m_{vv} - \lambda \, m_{x \theta}- \kappa_s \,m_{xx}
- G \, m_{x v}+ \sum_{\beta=1}^{N_i} \gamma_\beta \, \mu_\beta \,
m_{x z_\alpha} \nonumber \\
\frac{d m_{x \theta}}{d t} & = & m_{v \theta} - \lambda \, m_{x \theta}
+ m_{xv} \nonumber \\
\frac{d m_{x z_\alpha}}{d t} & = &  m_{v z_\alpha}- \mu_\alpha \, m_{x z_\alpha}+ m_{xv} \nonumber \\
\frac{d m_{vv}}{d t} & = & - 2 \, \lambda \, m_{v \theta}- 2 \, \kappa_s
\, m_{x v} -
2 \, G \, m_{vv}+ 2 \, \sum_{\beta=1}^{N_i}
\gamma_\beta \, \mu_\beta \,m_{v z_\beta}
+ 2 \, \sum_{\beta=1}^{N_i} d_\beta^2   \label{eqi_9} \\
\frac{d m_{v \theta}}{d t} & = & -\lambda \, m_{\theta \theta}
-\kappa_s \, m_{x \theta} - G \, m_{v \theta} + \sum_{\beta=1}^{N_i}
\gamma_\beta \, \mu_\beta
\, m_{\theta z_\beta} - \lambda  \, m_{v \theta}+ m_{vv} \nonumber \\
\frac{d m_{v z_\alpha}}{d t} & = &  -  \lambda \, m_{\theta z_\alpha} -
\kappa_s \, m_{x z_\alpha} - G \, m_{v z_\alpha}+
\sum_{\beta=1}^{N_i} \gamma_\beta \, \mu_\beta \, m_{z_\alpha z_\beta}
- \mu_\alpha \, m_{v z_\alpha}+ m_{vv}+ 2 \, c_\alpha \, d_\alpha
\nonumber \\
\frac{d m_{\theta \theta }}{d t} & = & - 2 \, \lambda \, m_{\theta \theta}+
2 \, m_{v \theta} + 2 \nonumber \\
\frac{d m_{\theta z_\alpha}}{d t} & = & - \lambda \, m_{\theta z_\alpha}
+ m_{v z_\alpha} - \mu_\alpha \, m_{\theta z_\alpha}+ m_{v \theta}
\nonumber \\
\frac{d m_{z_\alpha z_\beta }}{d t} & = &
- (\mu_\alpha+\mu_\beta) \,
+m_{v z_\alpha} + m_{v z_\beta} +2 \, c_\alpha^2 \, \delta_{\alpha \beta}
\nonumber
\end{eqnarray}
If the kernel is non singular at $t=0$, ($k(t)$ is singular in
the Newtonian case and it recovers  a finite value in  viscoelastic
models) then, independently of any non-vanishing viscoelastic contribution, one
should observe at very short times a $t^3$-scaling as
in the Stokesian case,
\begin{equation}
m_{xx}(t) \sim t^3
\label{eqi_10}
\end{equation}
 assuming that
the medium is in equilibrium (see figure \ref{Fig_a1}).
However this ``very early'' scaling may occur for timescales
$t \ll 10^{-3}$
and therefore these scales might not be reached within the actual
experimental resolution.

In contrast, a robust
$5/2$-scaling is observed over several decades,
\begin{equation}
m_{xx}(t) \sim t^{5/2}
\label{eqi_11a}
\end{equation}
 and this
intermediate scaling is the initial scaling
occurring in the presence of a  strictly singular
Basset kernel. Figure \ref{Fig_a1} shows that viscoelastic constitutive equations
do not modify the early $t^3$-scaling if a fluid inertial  contribution
is present.

Next, turn back attention to the Newtonian case. The results are summarized in figures \ref{Fig3}-\ref{Fig4} in the main text,
providing the occurrence of an initial $t^3$-scaling,
followed by an intermediate $t^{5/2}$-scaling.

In order to analyze the origin of the $t^{5/2}$-scaling,
 consider the moment dynamics with a Stokesian
friction eq. (\ref{eqi_7}) starting from a medium at equilibrium.
In this case we can assume $m_{z_\alpha z_\beta}(t)=m_{z_\alpha z_\beta}^*
= c_\alpha^2/\mu_\alpha = 1/\gamma_\alpha \mu_\alpha$.
At short timescales, the leading-order dynamics for $m_{xx}(t)$ 
reduce
to the  equations (derived by eq. (\ref{eqi_7}) neglecting
the higher-order contributions),
\begin{eqnarray}
\frac{d m_{xx}}{d t} & = & 2 \, m_{x v} \nonumber \\
\frac{d m_{xv}}{d t} & = &  m_{v v} \nonumber \\
\frac{d m_{vv}}{d t} & = & 2 \,  \sum_{\alpha=1}^{N_i} \gamma_\alpha \, \mu_\alpha \, m_{v z_\alpha} + 2 G +2 \label{eqi_12} \\
\frac{d m_{v z_\alpha}}{d t} & = &- (1+ G + \mu_\alpha ) \, m_{v z_\alpha}
+ \sum_{\beta=1}^{N_i} \gamma_\beta  \, \mu_\beta \, m_{z_\alpha z_\beta}^*
+ 2 \, c_\alpha \, d_\alpha \nonumber
\end{eqnarray}
Since $c_\alpha d_\alpha =-1$ for any $\alpha=1,\dots,N_i$,
and $\sum_{\beta=1}^{N_i} \gamma_\beta  \, \mu_\beta \, m_{z_\alpha z_\beta}^*
=1$, the $m_{v z_\alpha}$-dynamics becomes
\begin{equation}
\frac{d m_{v z_\alpha}}{d t}  = - (1+ G + \mu_\alpha ) \, m_{v z_\alpha}
-1 \, ,
\label{eqi_13}
\end{equation}
and since $m_{v z_\alpha}(0)=0$, the solution of the latter equation is
\begin{equation}
m_{v z_\alpha}(t)= - \frac{1}{1+G+\mu_\alpha} \left [ 1- e^{-(1+G+ \mu_\alpha)
t} \right ] \, .
\label{eqi_14}
\end{equation}
Consequently the dynamics for $m_{vv}(t)$ reduces to
\begin{eqnarray}
\frac{1}{2} \, \frac{d m_{vv}}{d t}  = 1+ G
- \sum_{\alpha=1}^{N_i}\frac{\gamma_\alpha , \mu_\alpha}{1+G+\mu_\alpha} \left [ 1- e^{-(1+G+ \mu_\alpha) \,
t} \right ]
 =  \phi(t) \, .
\label{eqi_15}
\end{eqnarray}
Given the scaling of the function $\phi(t)$ with time, say $\phi(t) \sim t^{\kappa}$
for small $t$, the scaling of $m_{xx}(t)$ simply follows,
\begin{equation}
m_{xx}(t) \sim t^{3+\kappa}  \, .
\label{eqi_16}
\end{equation}
The function $\phi(t)$ is the sum of two terms: a constant
term $\phi_c$ and a time-dependent contribution
$\phi_t(t)$,
\begin{eqnarray}
\phi_c & = & 1+ G
- \sum_{\alpha=1}^{N_i}\frac{\gamma_\alpha , \mu_\alpha}{1+G+\mu_\alpha}
\nonumber \\
\phi_t(t) & = & \sum_{\alpha=1}^{N_i}\frac{\gamma_\alpha \, \mu_\alpha}{1+G+\mu_\alpha}  \,  e^{-(1+G+ \mu_\alpha) \, t}
\label{eqi_17}
\end{eqnarray}
\begin{figure}
\includegraphics[width=12cm]{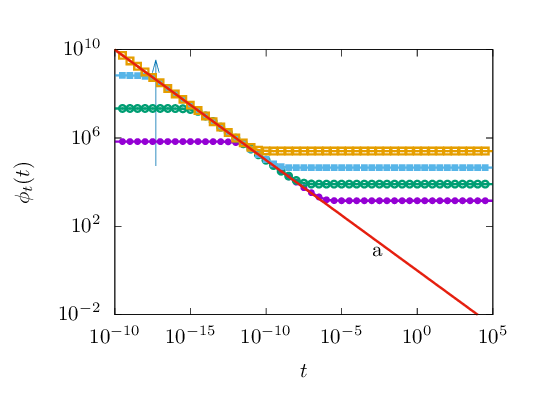}
\caption{$\phi_t(t)$ vs $t$ for different values of $\Gamma$.
The arrow indicates increasing values of $G=10^{12},\,10^{15},\,10^{18},10^{21}$.
Line (a) represents the curve $\phi_t(t)=1/\sqrt{t}$.}
\label{Fig_a2}
\end{figure}

In the case $G$ is finite, $\phi_c$ at short-time prevails
and therefore the exponent $\kappa=0$. Therefore,
the other case of interest is when $G$ is diverging. In this
case, if $\mu_\alpha > G$ (still in the limit for $G \rightarrow \infty$),
we have
\begin{eqnarray}
\phi_t(t)    =    \sum_{\alpha=1}^{N_i}\frac{\gamma_\alpha \, \mu_\alpha}{1+G+\mu_\alpha}  \,  e^{-(1+G+ \mu_\alpha) \, t}
 \sim \sum_{\alpha=1}^{N_i} \gamma_\alpha \, e^{-\mu_\alpha \, t} =
k(t)
\label{eqi_18}
\end{eqnarray}
Therefore, in the Basset case, $k(t)=\beta/\sqrt{t}$, $\kappa=-1/2$,
leading to the $5/2$-scaling of $m_{xx}(t)$.
The scaling of $\phi_t(t)$ vs $t$ is depicted in figure \ref{Fig_a2}.
As $\Gamma$ increases the scaling  $\phi_t(t)\sim 1/\sqrt{t}$  occurs
at short times, and in the limit as $\Gamma \rightarrow \infty$, this
becomes the initial scaling of the particle dispersion dynamics.

\subsubsection{Correlated fluctuational patterns}
\label{sec4.3}
A further generalization involves correlated
fluctuational patterns that have been introduced in \cite{gpp3},
in order to take into account the finite propagation velocity
of  hydrodynamic fluctuations in the solvent fluid.
The  physical motivation for this extension is the following:
the inclusion of acoustic effects accounting for the finite
propagation (bounded speed of sound) of density and compressible
stresses in a fluid determines  the disappearance of
the hydrodynamic added-mass effect.  For physical consistency,
one should also consider that  all the elementary stochastic
forcing entering the fluctuational patterns and acting
either on the velocity dynamics or on the dynamics of the
auxiliary variables should possess bounded propagation
velocity and finite correlation times.
This means to substitute, within the structure of
the correlation patterns,  the impulsive $\delta$-correlated forcings
with correlated stochastic processes
properly normalized in order to recover, in the limit of vanishing correlation
times, the  results obtained in the impulsively correlated case.

The model  with correlated stochastic forcings so constructed
still satisfies
the Stokes-Einstein global fluctuation-dissipation relation,
connecting particle diffusivity to its friction factor.
In the case of 
 eqs. (\ref{eqi_1}), the simplest correlated fluctuational pattern, characterized
by a single relaxation time $1/\nu$, with $\nu>0$ and bounded, is expressed by
\begin{eqnarray}
\frac{d x(t)}{d t} & = & v(t) \nonumber \\
\frac{ d v(t)}{d t} & = & - v(t) - \kappa_s \, x(t)- G \, v(t) +
\sum_{\alpha=1}^{N_i} \gamma_\alpha \, \mu_\alpha \, z_\alpha(t)
+ \sqrt{2} \,  \sum_{\alpha=1}^{N_i} d_\alpha \, q_\alpha(t) + \sqrt{2} \, q(t)  \nonumber \\
\frac{d z_\alpha(t)}{d t} & = & - \mu_\alpha \, z_\alpha(t) + v(t) + \sqrt{2}
\, c_\alpha\, q_\alpha(t) \label{eqco_1} \\
\frac{d q(t)}{d t}& = &- \nu \, \left ( q(t)-\xi(t) \right ) \nonumber \\
\frac{d q_\alpha(t)}{d t} & = & \nu \, \left ( q_\alpha(t)-\xi_\alpha(t)  \right ) \nonumber
\end{eqnarray}
with the same meaning for $\xi(t)$, $\xi_\alpha(t)$ as in eq. (\ref{eqi_1}) and of
the coefficients $c_\alpha$, $d_\alpha$, $\alpha=1,\dots,N_i$ as in eq. (\ref{eqi_2}).
The newly introduced processes $q(t)$, $q_\alpha(t)$, $\alpha=1,\dots,N_i$,
represent
correlated stochastic forcings
 possessing finite correlation rate $\nu$.
For $\nu \rightarrow \infty$, $q(t) \rightarrow \xi(t)$, $q_\alpha(t)
\rightarrow \xi_\alpha(t)$,
$\alpha=1,\dots,N_i$, with the convergence  to be interpreted in a weak distributional way, {one recovers eqs. (\ref{eqi_1}).

Also in this case, consider the preparation $x_0=v_0=0$ while the other degrees of freedom,
i.e. $z_\alpha(t)$, $q(t)$, $q_\alpha(t)$, $\alpha=1,\dots,N_i$ accounting for the
thermo-hydromechanic state of the solvent fluid, are in equilibrium.

Performing  second-order moment analysis as in the case of eq. (\ref{eqi_1}) -
the expression for the equations of the moment dynamics is not  reported for
the sake of  brevity, as conceptually analogous to  eqs. (\ref{eqi_7}) - the
initial scaling of $m_{xx}(t)$ starting from the Z-preparation is

\begin{equation}
m_{xx}(t) \sim t^4
\label{eqco_2}
\end{equation}
This phenomenon is depicted in figure \ref{Fig_a3}.
The  initial scaling eq. (\ref{eqco_2}) is followed
by a crossover at intermediate times to the inertially-controlled
scaling $m_{xx}(t) \sim t^{5/2}$. In fact, as the value of
$\Gamma$ increases, a series of crossovers can be detected:
from the early $t^4$-scaling, to an intermediate $t^3$-scaling
controlled by the finite value of $\Gamma$ and to a subsequent
$t^{5/2}$-scaling associated with the $1/\sqrt{t}$-behavior
of the fluid inertial kernel. This is shown in figure \ref{Fig_a4}.

\begin{figure}
\includegraphics[width=12cm]{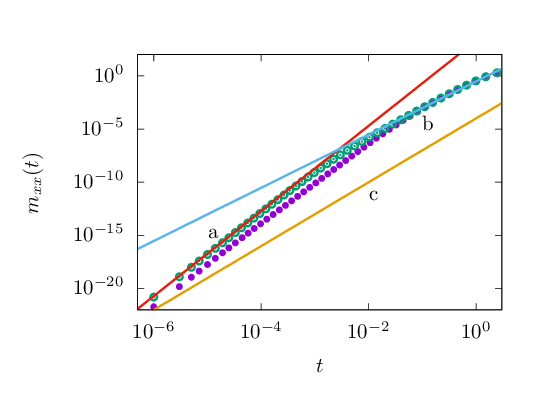}
\caption{$m_{xx}(t)$ vs $t$ (symbols)
for the correlated pattern defined by eq. (\ref{eqco_1})
at $\beta=\kappa_s=0.1$, $\Gamma=10^5$, and for different values of
$\nu$. Symbols ($\circ$) refer to $\nu=10^3$, ($\bullet$) to $\nu=10^2$.
Line (a) corresponds to the scaling $m_{xx}(t) \sim t^4$,
line (b) to the Basset scaling $m_{xx}(t) \sim t^{5/2}$, line (c)
to $m_{xx}(t) \sim t^3$.}
\label{Fig_a3}
\end{figure}

The $t^4$-initial scaling is generic for any correlated
pattern. To give an example, if one considers the
correlated pattern associated with the Einstein-Langevin
model eq. (\ref{eq4_8}) in a harmonic trap, namely,
\begin{eqnarray}
\frac{d x(t)}{d t} & = & v(t) \nonumber \\
\frac{d v(t)}{d t} & = & - v(t) - \kappa_s \, x(t) + \sqrt{2} \, q(t)
\label{eqco_3} \\
\frac{d q(t)}{d t} & = & - \nu \, \left ( q(t)-\xi(t) \right ) \nonumber
\end{eqnarray}
characterized by the moment equations
\begin{eqnarray}
\frac{d m_{xv}}{d t} & = & 2 \, m_{xv} \nonumber \\
\frac{d m_{xv}}{d t} & = & m_{vv}- m_{xv} -\kappa_s \, m_{xx}+\sqrt{2}\,m_{xq} \nonumber \\
\frac{d m_{xq}}{d t} & = & m_{vq}-\nu \, m_{xq} \label{eqco_3a} \\
\frac{d m_{vv}}{d t} & = & -2\, m_{vv}- 2 \, \kappa_s \,m_{xv}+ 2 \, \sqrt{2} \, m_{vq} \nonumber \\
\frac{d m_{vq}}{d t} & = & -m_{vq}- \kappa_s \, m_{xq} + \sqrt{2}  \, m_{qq} - \nu \,
m_{v q} \nonumber \\
\frac{d m_{qq}}{d t} &  = & - 2 \, \nu \, m_{qq} +\nu^2  \, ,\nonumber
\end{eqnarray}
then one obtains in the Z-preparation ($x_0=v_0=0$),
\begin{equation}
m_{vv}(t) \simeq \nu \, t^2 \, , \qquad m_{xx}(t) = \frac{\nu \, t^4}{4}
\label{eqco_4}
\end{equation}

\begin{figure}
\includegraphics[width=12cm]{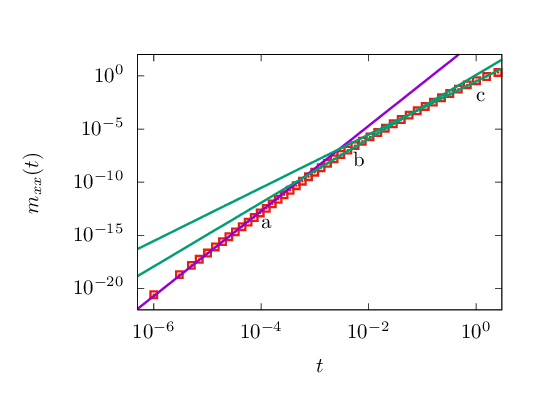}
\caption{$m_{xx}(t)$ vs $t$ (symbols)
for the correlated pattern defined by eq. (\ref{eqco_1})
at $\beta=\kappa_s=0.1$, $\Gamma=10^6$,
$\nu=10^3$, ($\bullet$).
Line (a) corresponds to the scaling $m_{xx}(t) \sim t^4$,
line (b) to the Basset scaling $m_{xx}(t) \sim t^{3}$, line (c)
to $m_{xx}(t) \sim t^{5/2}$.}
\label{Fig_a4}
\end{figure}

 \subsubsection{Anomalous dissipative and anomalous diffusive behavior}
\label{sec4.4}
In this paragraph we develop the modal representation
of Brownian motion in a complex fluid possessing a dissipative
kernel $h(t)$ admitting a long-term power-law behavior,
$h(t) \sim t^{-\omega}$ with $\omega \in (0,1)$, 
so that the   long-term dynamics are anomalous, in the
absence of inertial effects, i.e.,
$k(t)=0$.

Assuming for $h(t)= g_\omega(t,\Gamma,N_d)= \sum_{i=1}^{N_d}
h_i\, e^{-\lambda_i \, t}$, where $g_\omega(t,\Gamma,N_d)$ is
defined by eq.  (\ref{eqi_3}), and neglecting the
presence of the trap (that is immaterial in the short-time analysis),
the modal representation of the Brownian dynamics reads \cite{gpp1}
\begin{eqnarray}
\frac{d x(t)}{d t } & = & v(t) \nonumber \\
\frac{d v(t)}{d t}  & =  & - \sum_{i=1}^{N_d} h_i \, \theta_i(t)
\label{eq_d1} \\
\frac{d \theta_i(t)}{d t} &= & -\lambda_i \, \theta_i(t)+ v(t)
+ \sqrt{2} \, b_i \, \xi_i(t)
\nonumber
\end{eqnarray}
where $\xi_i(t)=d w_i(t)/d t$, $i=1,\dots,N_d$ are distributional
derivatives of independent Wiener processes $w_i(t)$, 
and $b_i=\sqrt{\lambda_i/h_i}$, $i=1,\dots,N_d$,  in
order to fulfill the Kubo fluctuation-dissipation relation of
the first kind.
At equilibrium,
\begin{equation}
\langle v^2 \rangle_{\rm eq}=1 \, , \quad
\langle \theta_i \, \theta_j \rangle_{\rm eq}= \frac{\delta_{ij}}{h_i}
\, , \quad \langle v \, \theta_i \rangle_{\rm eq}=0
\label{eq_d2}
\end{equation}
The Fokker-Planck equation for the probability
density $p(x,v,\boldsymbol{\theta},t)$ associated
with eq. (\ref{eq_d1}), where $\boldsymbol{\theta}=(\theta_1, \dots,
\theta_{N_d})$, reads
\begin{equation}
\frac{\partial p}{\partial t}  =  - v \, \frac{\partial p}{\partial x}
+\sum_{i=1}^{N_d} h_i \, \theta_i \, \frac{\partial p}{\partial v}
+ \sum_{i=1}^{N_d} \lambda_i \, \frac{\partial (\theta_i \, p)}{\partial 
\theta_i} - v  \, \sum_{i=1}^{N_d} \frac{\partial p}{\partial \theta_i}
+ \sum_{i=1}^{N_d} b_i^2 \, \frac{\partial^2 p}{\partial \theta_i^2}
\label{eq_d3}
\end{equation}
from which the dynamics of the second-order moments follows
\begin{eqnarray}
\frac{d m_{xx}}{d t} & = & 2 \, m_{xv} \nonumber \\
\frac{d m_{xv}}{d t} & = & m_vv - \sum_{i=1}^{N_d} h_i \, m_{x \theta_i} 
\nonumber \\
\frac{d m_{x \theta_i}}{d t} & = & m_{v \theta_i}-\lambda_i \, m_{x \theta_i}
+ m_{x v} \nonumber \\
\frac{d m_{vv}}{d t} & = & - 2 \, \sum_{i=1}^{N_d} h_i \, m_{v \theta_i}
\label{eq_d4} \\
\frac{d m_{v \theta_i}}{d t} & =  &- \sum_{j=1}^{N_d} h_j \, m_{\theta_i \theta_j}
-\lambda_i \, m_{v \theta_i}+ m_{vv} \nonumber \\
\frac{d m_{\theta_i \theta_j}}{d t} & = &  -(\lambda_i + \lambda_j)
\, m_{\theta_i \theta_j} + m_{v \theta_i}+ m_{v \theta_j}
+2  \, b_i^2  \delta_{ij}
\nonumber
\end{eqnarray}
Due to the regularity of the velocity fluctuations,
induced by the viscoelastic  dynamics and by the boundedness
of $h(0)$, the short-time scaling for
a Z-preparation  where $m_{xx}(0)=m_{xv}(0)=m_{x \theta_i}(0)= m_{vv}(0)= m_{v \theta_i}(0)=0$ and $m_{\theta_i \theta_j}=\langle \theta_i \theta_j \rangle_{\rm eq}$, provides a $t^4$-scaling for $m_{xx}(t)$.

\begin{figure}
\includegraphics[width=10cm]{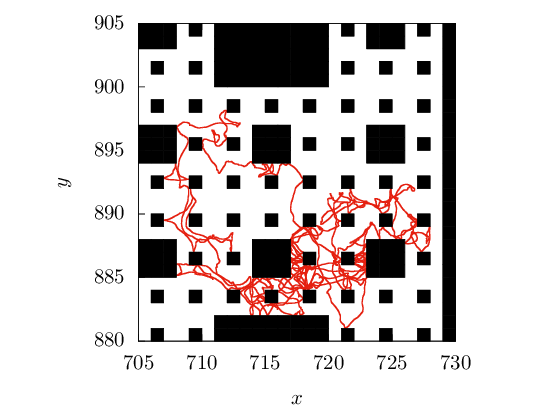}
\caption{Portion of an orbit of a Brownian particle
in a Sierpinski carpet subjeted exclusively to
Stokesian friction and thermal fluctuations.} 
\label{Fig_a5}
\end{figure}

The short-time analysis developed  in the  article applies also
in the case  of anomalous diffusion processes induced by
the constrained geometry of the medium, as in the case
of a physical fractal medium \cite{havlin}.
Figure \ref{Fig_a5} depicts a portion of an orbit
 of a Brownian particle in a two-dimendional Sierpinski carpet (having
fractal dimension $d_f=\log 8/\log 3 \approx 1.893$)
characterized by a  cell length $L_c$ of free motion
($L_c=1$ in the present analysis).

\begin{figure}
\includegraphics[width=12cm]{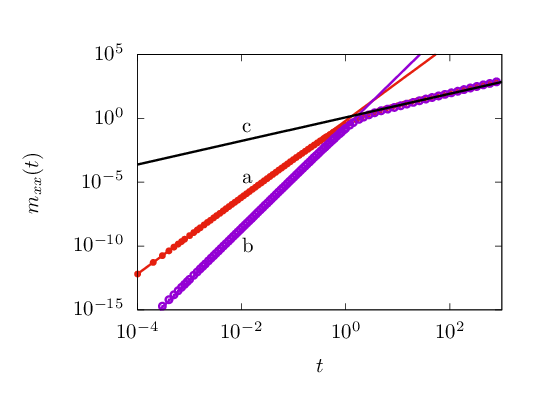}
\caption{ $m_{xx}(t)$ vs $t$ for Brownian
dynamics on a Sierpinski carpet. Symbols
represents the results of stochastic simulations:
($\bullet$) for a Stokesian dynamics, ($\circ$) for
the motion in a Maxwell fluid characterized
by a relaxation rate $\lambda=1$.
Line (a) corresponds to the scaling $m_{xx}(t) \sim t^3$,
line (b) to $m_{xx}(t) \sim t^4$, line (c) to
the asymptotic anomalous saling $m_{xx}(t) \sim
t^{2/d_w}$, with $d_w=2.139$.}
\label{Fig_a6}
\end{figure} 
Figure \ref{Fig_a6} shows the temporal evolution
of $m_{xx}(t)$ starting from a Z-preparation, obtained
from stochastic simulations of an ensemble of $N_p=10^5$
particles, in the case of a Stokesian dynamics
($h(t)=\delta (t)$, $k(t)=0$),
and for a Maxwell fluid ($h(t)= \lambda \, e^{-\lambda \, t}$,
$k(t)=0$) characterized by a relaxation rate $\lambda=1$.
The Langevin equations have been integrated numerically
via an Euler-Langevin algorithm with a step size
$h_h=10^{-3}$, enforcing reflecting boundary
onditions at the solid walls of the carpet structure.
As expeted from the analysis developed in the text,
the short-time  behavior is characterized
by the scaling $m_{xx}(t) \sim t^3$ in the Stokesian case,
and $m_{xx}(t) \sim t^4$ for a Maxwell fluid.
The long-term regime is characterized by an anomalous 
subdiffusive scaling $m_{xx}(t) \sim t^{2/d_w}$ \cite{havlin}
where $d_w$ is the walk dimension attaining, for
the considered Sierpinski structure, the value
$d_w=2.139 \pm 0.05$ \cite{sierpinski}.

\end{document}